\documentclass[12pt,a4paper,twoside]{article}

\usepackage[dvips]{graphics}
\usepackage{epsfig}
\usepackage{cite}
\usepackage{rotating}

\def\Journal#1#2#3#4{{#1} {\bf #2} (#3) #4}

\def\NIM{\em Nucl. Inst. Meth.}
\def\NIMA{{\em Nucl. Inst. Meth.} A}
\def\NPB{{\em Nucl. Phys.} B}

\def\PLB{{\em Phys. Lett.}  B}
\def\PL{\em Phys. Lett.}
\def\PRL{\em Phys. Rev. Lett.}
\def\PRD{{\em Phys. Rev.} D}
\def\PR{\em Phys. Rev.}

\def\EPJC{{\em Eur. Phys. J.} C}

\def\CPC{\em Comp. Phys. Comm.}
\def\ZFP{{\em Z.\ Phys.} C }

\def\JHEP{\em JHEP}
\def\etal{{\it et al.}}

\def\NPBPS{\em Nucl.\ Phys.\ B (Proc.\ Suppl.)}
\def\NPPS{\em Nucl.\ Phys.\ Proc.\ Suppl.}

\def\be{\begin{equation}}
\def\ee{\end{equation}}
\def\bea{\begin{eqnarray}}
\def\eea{\end{eqnarray}}

\newcommand{\result} {\ensuremath{\asmz=0.1182\pm0.0003(\mathrm{stat.})\pm0.0015(\mathrm{exp.})\pm0.0011(\mathrm{had.})\pm0.0012(\mathrm{scale})\pm0.0013(\mathrm{mass})}}
\newcommand{\restot} {\ensuremath{\asmz=0.1182\pm0.0025(\mathrm{total~error})}}
\newcommand{\resultxmufree} {\ensuremath{\asmz=0.1187\pm0.0005(\mathrm{stat.})\pm0.0018(\mathrm{exp.})\pm0.0011(\mathrm{had.})\pm0.0012(\mathrm{mass})}}
\newcommand{\rxmufree} {\ensuremath{\xmuopt=0.672\pm0.033(\mathrm{stat.})}}
\newcommand{\resultxmuasmin} {\ensuremath{\asmz=0.1182\pm0.0003(\mathrm{stat.})\pm0.0014(\mathrm{exp.})\pm0.0012(\mathrm{had.})\pm0.0013(\mathrm{mass})}}
\newcommand{\rxmumin} {\ensuremath{\xmuopt=1.36}}
\newcommand{\resdpxmu}{\ensuremath{\asmz=0.1047\pm0.0014(\mathrm{stat.})\pm0.0037(\mathrm{exp.})\pm0.0061(\mathrm{had.})\pm0.0052(\mathrm{scale})}}
\newcommand{\restmxmu}{\ensuremath{\asmz=0.1318\pm0.0016(\mathrm{stat.})\pm0.0056(\mathrm{exp.})\pm0.0023(\mathrm{had.})\pm0.0111(\mathrm{scale})}}
\newcommand{\resdp}{\ensuremath{\asmz=0.0987\pm0.0011(\mathrm{stat.})\pm0.0044(\mathrm{exp.})\pm0.0022(\mathrm{had.})}}
\newcommand{\restm}{\ensuremath{\asmz=0.1100\pm0.0011(\mathrm{stat.})\pm0.0045(\mathrm{exp.})\pm0.0032(\mathrm{had.})}}
\newcommand{\dxmufree} {\ensuremath{\xmuopt=0.0220\pm0.0002(\mathrm{stat.})}}
\newcommand{\txmufree} {\ensuremath{\xmuopt=0.0089\pm0.0001(\mathrm{stat.})}}
\newcommand{\Opal}{\mbox{\rm OPAL}}
\newcommand{\Delphi}{\mbox{\rm DELPHI}}
\newcommand{\Aleph}{\mbox{\rm ALEPH}}

\newcommand{\qqbar}     {\ensuremath{\mathrm{q\bar{q}}}}

\newcommand{\qqbargg}     {\ensuremath{\mathrm{q\bar{q}gg}}}
\newcommand{\qqbarqqbar}     {\ensuremath{\mathrm{q\bar{q}q\bar{q}}}}

\newcommand{\lnu}               {\ensuremath{\ell\nu}}

\newcommand{\epem}              {\ensuremath{\mathrm{e^+e^-}}}

\newcommand{\as}                {\ensuremath{\alpha_\mathrm{S}}}

\newcommand{\assq}    {\ensuremath{\alpha_\mathrm{S}^{\mathrm{2}}}}
\newcommand{\ascu}    {\ensuremath{\alpha_\mathrm{S}^{\mathrm{3}}}}

\newcommand{\asmz}              {\ensuremath{\alpha_\mathrm{S}(M_{\mathrm{Z^0}})}}

\newcommand{\zzero}     {\ensuremath{\mathrm{Z^0}}}

\newcommand{\mz}                {\ensuremath{M_{\mathrm{Z^0}}}}

\newcommand{\tma}               {\ensuremath{T_{\mathrm{major}}}}
\newcommand{\tmi}               {\ensuremath{T_{\mathrm{min}}}}

\newcommand{\chisqd}    {\ensuremath{\chi^2/\mathrm{d.o.f.}}}
\newcommand{\xmu}               {\ensuremath{x_{\mu}}}

\newcommand{\xmuopt}               {\ensuremath{x_{\mu}^\mathrm{opt}}}
\newcommand{\xmumin}               {\ensuremath{x_{\mu}^\mathrm{min}}}

\newcommand{\ycut}              {\ensuremath{y_{\mathrm{cut}}}}

\newcommand{\rs}                {\ensuremath{\sqrt{s}}}

\newcommand{\gamgam}    {\ensuremath{\gamma^{*}\gamma^{*}}}

\newcommand{\ww}                {\ensuremath{\mathrm{W^+W^-}}}

\newcommand{\invpb}     {\ensuremath{\mathrm{pb}^{-1}}}

\newcommand{\bm}[1]     {\mbox{\boldmath\ensuremath{#1}}}

\newcommand{\py}                {PYTHIA}
\newcommand{\hw}                {HERWIG}
\newcommand{\ar}                {ARIADNE}
\newcommand{\jt}                {JETSET}

\newcommand{\kw}                {KORALW}
\newcommand{\grc}  		{GRC4f}

\newcommand{\debr}    {Debrecen 2.0}

\begin{document}

\begin{titlepage}

\centerline{{\large EUROPEAN ORGANIZATION FOR NUCLEAR RESEARCH}} \bigskip

\begin{flushright}
PH-EP/2005-057\\
25th November 2005
\end{flushright}

\bigskip\bigskip\bigskip

\begin{center}\textbf{\Large Measurement of the Strong Coupling \boldmath{\as} from
Four-Jet Observables in \epem\ Annihilation}\end{center}{\Large \par}

\bigskip

\begin{center}{\Large The \Opal\ Collaboration}\end{center}\par

\bigskip

\begin{abstract}
Data from \epem\ annihilation into hadrons at centre-of-mass energies 
between 91~GeV and 209~GeV collected with the \Opal\ detector at LEP, 
are used to study the four-jet rate as 
a function of the Durham algorithm resolution parameter \ycut. The four-jet
rate is compared to next-to-leading order calculations that include the resummation
of large logarithms. The strong coupling measured from the four-jet rate is 
\begin{center}
\result, \\
\end{center}
in agreement with the world average. Next-to-leading order fits to 
the $D$-parameter and thrust minor event-shape observables are also 
performed for the first time.
We find consistent results, but with significantly larger theoretical uncertainties.
\end{abstract}

\bigskip\bigskip\bigskip\bigskip\bigskip

{\large \par}

\vfill
\end{titlepage}

\begin{center}{\Large        The OPAL Collaboration
}\end{center}\bigskip
\begin{center}{
%begin authorlist PLEASE DO NOT DELETE THIS COMMENT
G.\thinspace Abbiendi$^{  2}$,
C.\thinspace Ainsley$^{  5}$,
P.F.\thinspace {\AA}kesson$^{  3,  y}$,
G.\thinspace Alexander$^{ 22}$,
G.\thinspace Anagnostou$^{  1}$,
K.J.\thinspace Anderson$^{  9}$,
S.\thinspace Asai$^{ 23}$,
D.\thinspace Axen$^{ 27}$,
I.\thinspace Bailey$^{ 26}$,
E.\thinspace Barberio$^{  8,   p}$
% ...
%end authorlist PLEASE DO NOT DELETE THIS COMMENT
}\end{center}\bigskip
\bigskip
%begin institutes
$^{  1}$School of Physics and Astronomy, University of Birmingham,
Birmingham B15 2TT, UK
\newline
$^{  2}$Dipartimento di Fisica dell' Universit\`a di Bologna and INFN,
I-40126 Bologna, Italy
\newline
$^{  3}$Physikalisches Institut, Universit\"at Bonn,
D-53115 Bonn, Germany
\newline
% ...
$^{  5}$Cavendish Laboratory, Cambridge CB3 0HE, UK
\newline
% ...
$^{  8}$CERN, European Organisation for Nuclear Research,
CH-1211 Geneva 23, Switzerland
\newline
$^{  9}$Enrico Fermi Institute and Department of Physics,
University of Chicago, Chicago IL 60637, USA
\newline
% ...
$^{ 22}$Department of Physics and Astronomy, Tel Aviv University,
Tel Aviv 69978, Israel
\newline
$^{ 23}$International Centre for Elementary Particle Physics and
Department of Physics, University of Tokyo, Tokyo 113-0033, and
Kobe University, Kobe 657-8501, Japan
\newline
% ...
$^{ 26}$University of Victoria, Department of Physics, P O Box 3055,
Victoria BC V8W 3P6, Canada
\newline
$^{ 27}$University of British Columbia, Department of Physics,
Vancouver BC V6T 1Z1, Canada
\newline
% ...
$^{  p}$ now at The University of Melbourne, Victoria, Australia
\newline
% ...
$^{  y}$ now at CERN
\newline
% ...
%end notes
\section{Introduction}
\label{intro}
The annihilation of electrons and positrons into hadrons allows
a precise test of Quantum Chromodynamics (QCD). Many observables
have been devised which provide a convenient way of characterizing
the main features of such events.
Multi-jet rates are predicted in perturbation theory as functions
of the jet-resolution parameter, with one free parameter, the
strong coupling \as.
Events with four quarks in the final state, \qqbarqqbar, or two quarks and two 
gluons, \qqbargg, may lead to events with four-jet structure.
In leading order perturbation theory, the rate of four-jet events
in \epem\ annihilation is predicted to be proportional to \assq.
Thus, the strong coupling is measured by determining the 
four-jet production rate in hadronic events and fitting
the theoretical prediction to the data.\par
Calculations beyond leading order are made possible by theoretical
progress achieved during the last few years. 
For multi-jet rates as well as numerous event-shape 
distributions with perturbative expansions
starting at $\cal{O}(\as)$, matched next-to-leading order (NLO) and next-to-leading
logarithmic approximations (NLLA) satisfactorily
describe data over large kinematically allowed regions at many centre-of-mass
energies~\cite{CTWT,CDOTW,DSjets,NTqcd98}. \par
Measuring the four-jet rate is not the only way to determine
the strong coupling; various event-shape observables
sensitive to the presence of four partons in the final state
may also be studied.
In addition to measuring the four-jet rates, we also studied
two event-shape observables which are sensitive to the four 
jet structure of the final state, the $D$-parameter~\cite{dpar} and the thrust 
minor (\tmi) ~\cite{tmin}. For these observables the NLLA predictions 
are only available in the nearly planar limit.
Furthermore, the NLO corrections are rather large,
which indicates that the neglected higher order terms in the 
perturbative prediction are important~\cite{dparbad,campbell}
and lead to a large uncertainty in the determination
of the strong coupling. \par
In this analysis we use data collected by \Opal\ at LEP
at 13 centre-of-mass energies covering the full LEP energy range of
91$-$209~GeV.
A previous publication by the \Opal\ Collaboration performed 
a simultaneous measurement of \as\ and the QCD colour factors with
data taken at 91~GeV~\cite{OPALCF}. The same theoretical predictions
for the four-jet rate are used for this analysis, with the colour factors 
$C_{\mathrm A}$ and $C_{\mathrm F}$ set to the values expected from the 
QCD SU(3) symmetry group, $C_{\mathrm A}=3$ and $C_{\mathrm F}=4/3$.
More importantly, our analysis extends the energy range up to 209~GeV
and we also investigate the $D$-parameter and \tmi\ event-shape observables.
Similar results from other LEP collaborations can be found 
in~\cite{alephr4,delphir4}.\par
The OPAL Collaboration has also published a comprehensive
measurement of jet-rates in the energy range between 
91~and 209~GeV~\cite{prmike} and the distributions of 
the $D$-parameter and \tmi\ amongst many other event
shape observables~\cite{pr404}. The data used in the present 
paper are identical to those presented in~\cite{prmike,pr404}.  
However, the determinations of \as\ in~\cite{prmike,pr404} were 
based on the two-jet rate and event shapes sensitive to 
three-jet processes; in contrast the present paper
focuses on different calculations for observables sensitive to
four-jet production.
\section{Observables}
\label{theoryall}
\subsection{The Four-Jet Rate}
\label{theoryr4}
Jet algorithms are applied to cluster the large number
of particles of a hadronic event into a small number of jets,
reflecting the parton structure of the event. For this
analysis we use the Durham scheme~\cite{CDOTW}. Defining 
each particle initially to be a proto-jet, a resolution variable $y_{ij}$ 
is calculated for each pair of proto-jets $i$ and $j$:
\be
 y_{ij}=\frac{2\mathrm{min}(E_i^2,E_j^2)}{E_{\mathrm{vis}}^2}(1-\cos\theta_{ij}),
\ee
where $E_{i}$ and  $E_{j}$ are the energies of jets $i$ and $j$,
$\cos\theta_{ij}$ is the angle between them and $E_{\mathrm{vis}}$ is the sum
of the energies of the visible particles in the event (or the
partons in a theoretical calculation). 
If the smallest
value of $y_{ij}$ is less than a predefined value $\ycut$, the pair
is replaced by a new proto-jet with four-momentum
$p_{k}^\mu =  p_i^\mu + p_j^\mu$, and the clustering starts again. 
Clustering ends
when the smallest value of $y_{ij}$ is larger than $\ycut$, and the remaining
proto-jets are accepted and counted as final selected jets.\par
The NLO calculation predicts the four-jet 
rate $R_{4}$, which is the fraction of four-jet events, as a function
of \as\ and \ycut. The NLO prediction can be written in the following way~\cite{zoltan}:
\be
R_{4}(y_{\mathrm{cut}})=\frac{\sigma_{\mathrm{\mbox{\scriptsize{4-jet}}}}(y_{\mathrm{cut}})}{\sigma_{\mbox{\scriptsize tot}}} \\
   = \eta^{2}B_4(y_{\mathrm{cut}})+\eta^{3}[C_4(y_{\mathrm{cut}})+3/2(\beta_{0}\log{\xmu}
     -1)\ B_4(y_{\mathrm{cut}})]
\label{NLOcalc}        
\ee
where $\sigma_{\mathrm{\mbox{\scriptsize{4-jet}}}}$ is the cross-section
for the production of a hadronic event with four jets at fixed \ycut,
$\sigma_{\mathrm{\mbox{\scriptsize tot}}}$ the total hadronic cross-section,
$\eta = \as \ C_{\mathrm F} / 2\pi$, $\xmu=\mu/\sqrt{s}$ with 
$\mu$ being the renormalization scale, $\sqrt{s}$ the centre-of-mass energy, 
and $\beta_{0}=(11-2n_{f}/3)$ with $n_{f}$ the number of active 
flavours\footnote{In this analysis the number of active flavours is set to five.}.  
The coefficients $B_4$ and $C_4$ are obtained by
integrating the massless matrix elements for \epem\
annihilation into four- and five-parton final states, calculated
by the program~\debr~\cite{zoltan}\footnote{The Durham 
$y_{\mathrm{cut}}$ values are chosen to vary in the 
range between $0.00001$ and $0.3162$, similar 
to the study in Ref.~\cite{zoltan}.}. Eq.~(\ref{NLOcalc}) is used to predict the four-jet rate
as a function of $y_{\mathrm{cut}}$. The fixed-order perturbative
prediction is not reliable for small values of $y_{\mathrm{cut}}$ due
to terms of the form $\as^{n}\ln^{m}(y_{\mathrm{cut}})$ which enhance the higher order 
corrections. An all-order resummation of such terms (NLLA) is possible for
the Durham clustering algorithm ~\cite{CDOTW}.
The NLLA calculation is combined with the NLO-prediction using the 
so-called modified R-matching scheme~\cite{OPALCF}.
In this
scheme the terms proportional to
$\eta^2$ and $\eta^3$ are removed from the NLLA 
prediction $R^{\mathrm{NLLA}}$ and the difference is then added to the NLO
calculation $R_{4}$ from Eq.~\ref{NLOcalc} with the result:
\be
 R^{\mathrm{R-match}}=R^{\mathrm{NLLA}}+[\eta^{2}(B_{4}-B^{\mathrm{NLLA}})+\eta^{3}(C_{4}-C^{\mathrm{NLLA}}-3/2(B_{4}-B^{\mathrm{NLLA}}))],
\label{NLLA}
\ee
where $B^{\mathrm{NLLA}}$ and $C^{\mathrm{NLLA}}$ are the coefficients of the 
expansion of $R^{\mathrm{NLLA}}$, taking the same form as 
Eq.~(\ref{NLOcalc}), and the \xmu\ term and
\ycut\ dependence have been suppressed for clarity. \par
\subsection{Thrust minor (\boldmath{\tmi}) and \boldmath{$D$-parameter}}
\label{theorydp}
The properties of hadronic events may also be described by 
event-shape observables.  These are used to characterize the
distribution of particles in an event as ``pencil-like'',
planar, spherical, etc. 
The variable \tmi\ and the $D$-parameter are sensitive
to hadronic configurations with more than three partons 
in the final state. These
observables are defined as follows:
\begin{description}
\item[Thrust minor \bm{\tmi}:]
  The thrust is defined by the expression\cite{tmin}
  \begin{equation}
  T= \max_{\vec{n}}\left(\frac{\sum_i|\vec{p_i}\cdot\vec{n}|}
                    {\sum_i|\vec{p_i}|}\right)\;,
  \label{equ_thrust}
  \end{equation}
  where $\vec{p_i}$ is the three-momentum of particle $i$ and
  the summation runs over all particles.
  The thrust axis $\vec{n}_T$ is the direction $\vec{n}$ which
  maximizes the expression in parentheses. 
  In a further step the maximization 
  in Eq.~(\ref{equ_thrust}) is performed with
  all momenta projected onto the plane perpendicular to $\vec{n}_T$.	
  The resulting vector is called $\vec{n}_{\tma}$.
  At the last step the expression in parentheses in 
  Eq.~(\ref{equ_thrust})
  is evaluated again for the vector
  $\vec{n}_{\tmi}$ which is perpendicular to both $\vec{n}_T$ and 
  $\vec{n}_{\tma}$, this is \tmi.
\item[\bm{D}-parameter:]
  This observable is based on the linearized momentum tensor 
  \begin{equation}
    \Theta^{\alpha\beta}= \frac{\sum_i(p_i^{\alpha}p_i^{\beta})/|p_i|}
                               {\sum_i|p_i|}\;\;\;,
                           \;\;\;\alpha,\beta= 1,2,3\;,
  \end{equation}
  where the sum runs over particles, $i$, and $\alpha$ and $\beta$ denote 
  the Cartesian coordinates of the momentum vectors. 
  The three eigenvalues $\lambda_j$ of this tensor define $D$ through
\begin{equation}D=27\lambda_1\lambda_2\lambda_3 \;.\end{equation}
\end{description}
The NLO calculation predicts the event-shape observable 
 distributions
\be
 \frac{1}{\sigma_{\mbox{\scriptsize tot}}}\frac{d\sigma}{dO_{4}}
   = \eta^{2}B_{O_{4}}(O_{4})+\eta^{3}[C_{O_{4}}(O_{4}) + 3/2(\beta_{0}\log{\xmu}-1)\ B_{O_{4}}(O_{4})]
\ee
with $O_{4}$ being the $D$-parameter or \tmi. 
The coefficients $B_{O_{4}}$ and $C_{O_{4}}$ are obtained by
integrating the massless matrix elements for \epem\
annihilation into four-parton final states, calculated
by the program~\debr.
\section{Analysis Procedure}
Although the data used in the present paper are identical to those
presented in~\cite{prmike,pr404}, for the sake of  completeness, and in
order to present a coherent discussion of the systematic
uncertainties, we give a brief account of the analysis procedure
below.
\subsection{The OPAL Detector}
\label{sec_detector}
A detailed description of the \Opal\
detector can be found in Ref.~\cite{OPALtech}. This analysis
relies on the reconstruction of charged particle trajectories
and on the measurement of energy depositions in the electromagnetic
calorimeters.
Tracking of charged particles was performed with the central detector,
which was located in a solenoidal magnet providing
an axial magnetic field of 0.435 T. The central detector contained
a silicon microvertex detector, and three drift chamber devices:
an inner vertex chamber, a large volume jet chamber and surrounding $z$-chambers
to measure the $z$-coordinate\footnote{In the \Opal\ right-handed
coordinate system the $x$-axis pointed towards the centre of the LEP
ring, the $y$-axis pointed approximately upwards and the $z$-axis pointed
in the direction of the electron beam. The polar angle $\theta$ and the
azimuthal angle $\phi$ are defined with respect to $z$ and $x$,
respectively, while $r$ is the distance from the $z$-axis.}. \
Most of the tracking information was obtained from the jet chamber,
which provided up to 159 measured space
points per track, and almost 100\,\% tracking efficiency in the region
$|\cos \theta|<0.98$. 
Electromagnetic energy was measured by lead glass calorimeters
surrounding the magnet coil, separated into a barrel ($|\cos
\theta|<0.82$) and two end-cap ($0.81<|\cos \theta|<0.98$) sections. The
electromagnetic calorimeter consisted of 11\,704 lead glass blocks with a
depth of 24.6 radiation lengths in the barrel and more than 22
radiation lengths in the end-caps. 
\subsection{Data Samples}
The data used in this analysis were collected by the \Opal\ detector 
between 1995 and 2000 and 
correspond to 14.5 \invpb\ of 91 GeV data, 10.4 \invpb\ of LEP1.5 data with
centre-of-mass energies of 130 GeV and 136 GeV and 706.4 \invpb\ of LEP2 data with
centre-of-mass energies ranging from 161 to 209~GeV. The highest energy runs
at average centre-of-mass energies between 192 and 207~GeV have an 
energy spread of about 1-2~GeV and we grouped these into six main energy points.
The data at 91~GeV were taken during calibration runs at the \zzero\ peak at 
the beginning of each year and during data-taking in 1996-2000.
The breakdown of the data samples, the mean centre-of-mass energies, the energy range,
the data-taking years and the collected luminosities are
given in Table~\ref{lumi}.
\begin{table}[h]
\begin{center}
\begin{tabular}{|r|r|r|r|r|r|} \hline
average& energy & year & luminosity  & events & events  \\
energy in GeV& range in GeV& & (\invpb) &selected & predicted  \\
\hline
91.3  & 91.0--91.5 &1996-2000 & 14.7 & 395695 &  $-$ \\
\hline
130.1 & 129.9--130.2&1995, 1997 &5.31 & 318 & 368.4\\
136.1 & 136.0--136.3&1995, 1997 &5.95 & 312 & 329.7\\
\hline
161.3 & 161.2--161.6&1996 &10.1 & 281 & 275.3 \\
172.1 & 170.2--172.5&1996 &10.4 & 218 & 232.2\\
182.7 & 180.8--184.2&1997 &57.7 & 1077 & 1083.5\\
\hline
188.6 & 188.3--189.1&1998 &185.2 & 3086 & 3130.1\\
191.6 & 191.4--192.1&1999 &29.5 & 514 & 473.0\\
195.5 & 195.4--196.1&1999 &76.7 & 1137 & 1161.3\\
199.5 & 199.1--200.2&1999,~2000 &79.3 & 1090 & 1130.8 \\
201.6 & 201.3--202.1&1999,~2000 &37.8 & 519 & 526.5\\
204.9 & 202.5--205.5&2000 &82.0 & 1130 & 1089.6\\
206.6 & 205.5--208.9&2000 &138.8 & 1717 & 1804.1\\
\hline
\end{tabular}
\caption{
The average center-of-mass energy, the energy range, the year
of data taking and the integrated luminosity for each data
sample, together with the numbers of selected data
and number of events expected from Monte Carlo simulation.
The horizontal lines separate the four energy intervals.}
\label{lumi}
\end{center}
\end{table} 
For presentation purposes we combine the data above 91 GeV 
into three energy points~\cite{pr404}.
In the combination the individual data sets are weighted by the product
of their luminosity and signal cross-section. 
The LEP1.5 data samples provide an energy point at 133.3~GeV, 
while the LEP2 samples give energy points at 177.4~GeV and 197.0~GeV corresponding 
to the energy ranges from
161 to 184~GeV and from 188 to 209~GeV. 
\subsection{Monte Carlo Samples}
The same samples of Monte Carlo simulated events  as in the analysis 
of~\cite{pr404} are used to correct the 
data for experimental effects such as acceptance, resolution and 
backgrounds. The process $\epem \to \qqbar$ 
is  simulated using \jt\ 7.4\cite{jetset}  at $\sqrt{s}$=91.2~GeV, and 
using KK2f\cite{kk2f} with fragmentation performed by \py\ 6.125\cite{pythia}
at higher energies.
Corresponding samples using \hw\ 6.2\cite{herwig} are used for systematic checks. 
Electroweak four-fermion
background processes are simulated using \kw\ 1.42\cite{koralw} with \grc\cite{grc4f}
matrix elements and fragmentation performed by \py. 
The Monte Carlo samples generated at each energy point are processed through 
a full simulation of the \Opal\ detector~\cite{gopal} and reconstructed in the same
way as the data.
In addition, for comparisons with the corrected data
and to estimate fragmentation effects, large samples
of generator-level Monte Carlo events are employed, using
the parton shower models \py\ 6.158, \hw\ 6.2 and \ar\ 4.11 \cite{ariadne}. 
Each of these fragmentation 
models contains a number of tunable parameters; 
these were adjusted by tuning to previously published \Opal\ data at 
$\sqrt{s}\sim91$~GeV as described in Ref.~\cite{OPALPR141} for 
\py/\jt\ and in Ref.~\cite{OPALPR379} for \hw\ and \ar. 
\subsection{Selection of Events}
The selection of events for this analysis follows exactly
the study described in~\cite{pr404}. 
It consists of three main stages: the
identification of hadronic event candidates, the removal of events with a 
large amount of initial state radiation (ISR), and the removal of four-fermion
background events. \par
The selection of hadronic events is based on 
cuts on particle multiplicity mainly to remove leptonic final states and 
visible energy and longitudinal momentum
balance mainly to remove two-photon events, $\epem \to \epem \gamgam$, with
hadronic final states. \par
Standard criteria~\cite{pr404} are used to select well measured tracks and clusters 
of energy deposits in the calorimeter.
The number of good charged particle
tracks is required to be greater than six.
The polar angle of the thrust axis is required to 
satisfy $|\cos(\theta_{\mathrm T})|<0.9$ in order that
the events be well contained within the detector acceptance.\par
The effective \epem\ centre-of-mass energy after 
excluding all ISR photons, $\sqrt{s'}$, is
estimated for each selected event using the algorithm described in Ref.~\cite{sellep2}. At 
centre-of-mass energies of 130~GeV and above, we require that  $\sqrt{s}-\sqrt{s'}<10$~GeV
in order to select non-radiative events. \par
At energies above the \ww\ production threshold, electroweak four-fermion events,
especially those involving \qqbarqqbar\ final states, become a substantial
background. 
These are reduced by using the standard \Opal\ \ww\ likelihood-based selections~\cite{wwsel}.
At centre-of-mass energies of 161~GeV and above, the $\ww\to\qqbarqqbar\ $ likelihood 
is required to satisfy ${\cal L}_{\qqbarqqbar}<0.25$ and the $\ww \to \qqbar\lnu$ likelihood
is required to satisfy ${\cal L}_{\qqbar\lnu}<0.5$.\par
The numbers of selected candidate events obtained
after applying the selection cuts are given in Table~\ref{lumi}. The 
numbers are consistent with expectations based
on Monte Carlo simulation\footnote{The numbers of events 
expected on the basis  of Monte Carlo simulations are given 
in all cases except 
for 91~GeV; to perform an accurate prediction close to 
the \zzero\ peak would require a much more 
careful investigation of the beam energy and luminosity 
than is required for the present analysis.}.
After all cuts, the acceptance for
non-radiative signal events\footnote{Defined for this purpose as those 
fulfilling  $\sqrt{s}-\sqrt{s'}< 1$~GeV, with 
$\sqrt{s'}$ the effective \epem\ centre-of-mass energy after 
excluding all ISR photons.}  
ranges from $88.5\%$ at 91~GeV, where the loss
in acceptance is mostly due to the cut on the polar angle of the thrust axis, 
to $76.5\%$ at 207~GeV, where additional cuts on $\sqrt{s'}$ and four-fermion 
background rejection reduce the efficiency. The residual four-fermion 
background is negligible below 161~GeV, and increases from $2.1\%$ at 
161~GeV to $6.2\%$ at 207~GeV.
\subsection{Corrections to the data}
\label{detectorcorrection}
In order to mitigate the effects of double counting of energy in tracks
and calorimetry, a standard algorithm is adopted~\cite{pr404} which
associates charged particles with electromagnetic calorimeter clusters, and
subtracts the estimated contribution of the charged particles
from the calorimetry energy. All selected tracks, and the
electromagnetic calorimeter clusters remaining after this
procedure, 
are used in the evaluation of the distributions.
The resulting distributions with all selection cuts applied
are called detector-level distributions.\par
The expected number of remaining four-fermion background events,
$\zeta_{i}$, is
subtracted from the number of data events, $N_{i}$, for each 
data point of each distribution\footnote{A data point
is the center of a bin of a distribution for the $D$-parameter
and \tmi, while for the four-jet rate we have
points at discrete values of \ycut\ where the analysis is performed.}. 
The effects of detector acceptance and resolution
and of residual ISR are then accounted for at each data point.
Since in the present analysis the Monte Carlo
model gives a good description of the data, and 
migration between data points is small, this procedure is justified. \par
Two distributions are formed from Monte Carlo 
simulated signal events for each observable; the first, at the detector level,
treats the Monte Carlo events identically to the data, while the
second, at the hadron level, is computed using the true four-momenta
of the stable particles\footnote{ All charged and 
neutral particles with a lifetime longer
than $3\times 10^{-10}$~s are treated as stable.} in the event, and is restricted
to events whose $s'$ satisfied $\sqrt{s}-\sqrt{s'}< 1$~GeV. The Monte Carlo ratio
of the hadron level to the detector level for each data point, $\alpha_{i}$,
is used as a correction factor for the data, yielding the
corrected data point $\tilde{N_{i}}=\alpha_{i}(N_{i}-\zeta_{i})$.\par
The hadron-level distribution is then normalized to unity: 
$P_{i}= \tilde{N_{i}}/N$, where the sum $N=\sum_{k} \tilde{N_{k}}$
includes the underflow and overflow data points where appropriate. 
\subsection{Fit Procedure}
Our measurement of the strong coupling \as\ is based on 
fits of QCD predictions to the corrected distributions.
The theoretical predictions as described 
in Section~\ref{theoryall} provide distributions at the parton level.
The parton-level distributions are obtained from the partons 
after the parton shower, just before the hadronization. 
In order to confront the theory with the hadron-level data,
it is necessary to correct for hadronization
effects. This is done by calculating the
distributions at both the hadron and the parton level
using \py\ and, as a cross-check, with the \hw\ and \ar\ models. 
The theoretical prediction is then multiplied by the ratio of the 
hadron- and parton-level distributions.\par
A $\chi^{2}$-value for each energy point is calculated using 
the following formula
\be
 \chi^{2}=\sum_{i,j}^{n}(O_{4,i}-O_{4}(\as)_{i}^{\mathrm{theo}})(V_{ij}(O_{4}))^{-1}(O_{4,j}-O(\as)_{j}^{\mathrm{theo}}),
\ee 
with $i,j$ running over all data points in the fit range and  $O_{4}$ corresponding 
to either the four-jet rate $R_{4}$, the differential distribution \tmi\ or the 
$D$-parameter, while $V_{ij}(O_{4})$ is the covariance matrix.
The $\chi^{2}$ value is minimized with respect to \as\ for each
energy point separately. \par
The four-jet rate is an integrated distribution, while
the $D$-parameter and \tmi\ distributions are differential ones.
Therefore in the four-jet rate distribution a single event can 
contribute to  several $y_{\mathrm{cut}}$-data point
and for this reason the data points are correlated.
The complete covariance matrix $W_{ij}$ is 
determined from four-jet rate distributions calculated at the 
hadron level. Subsamples are built by choosing 1000 events
randomly out of the set of all generated Monte 
Carlo events. A single event may be included in several subsamples, 
but the impact on the final covariance matrix 
is expected to be small and, therefore, is neglected~\cite{bootstrap}.
For every energy point 1000 subsamples are built. 
The covariance matrix is then used
to determine the correlation matrix, 
$\rho_{ij}=W_{ij}/\tilde{\sigma_{i}}\tilde{\sigma_{j}}$, 
with $\tilde{\sigma_{i}}=\sqrt{W_{ii}}$. The covariance matrix $V_{ij}(R_{4})$ used in
the $\chi^{2}$ fit is then determined using the statistical error 
$\sigma_{i}$ of the data sample at data point $i$ and the correlation matrix 
$\rho_{ij}:V_{ij}(R_{4})=\rho_{ij}\sigma_{i}\sigma_{j}$. \par 
For the fit to the event-shape observables the covariance matrix
$V_{ij}(O_{4})$ for $P_i$ is computed by transforming the diagonal 
Poisson  covariance matrix for the uncorrected data points~$N_i$~\cite{pr404}:
\begin{equation}
V_{ij} = \sum_k \frac{\partial P_i}{\partial N_k}
                \frac{\partial P_j}{\partial N_k} N_k
       = \frac{1}{N^4} \sum_k \alpha_k^2 N_k
         \left(N\delta_{ik} - \widetilde{N}_i\right)
         \left(N\delta_{jk} - \widetilde{N}_j\right) \;.
\end{equation}
\subsection{Combination of Energy Points}
\label{combination}
For presentation purposes the data of several energy points are combined. 
The four-jet rate and event shape distributions measured at the 
different energy points are averaged using the products of 
luminosity and cross-section.
The fit results for \as, however, are combined using 
the procedure of Ref.~\cite{pr404}.\par
In brief the method is as follows. The \as\ measurements to be combined
are first evolved to the common scale of the combination, $Q_{0}=\sqrt{s_{0}}$, 
assuming the validity of QCD. The measurements are then combined using 
a weighted mean method, based on minimizing the $\chi^{2}$ between the combined values and the 
measurements. If the measured values evolved to a common scale 
$Q_{0}=\sqrt{s_{0}}$ are denoted 
$y_{i}$, with covariance matrix $V^{\prime}$, the combined values, \as($\sqrt{s_{0}}$),
are given by
\be
\as(\sqrt{s_{0}})=\sum_{i} w_{i} y_{i} \;\;\;\; \mathrm{where} \;\;\;\; w_{i}=\frac{\sum_{j}(V^{\prime~-1})_{ij}}{\sum_{j,k}(V^{\prime~-1})_{jk}},
\label{weight}
\ee
and $i,j$ enumerate the individual measurements.
Only experimental systematic  errors (assumed to be fully correlated, 
$V^{\prime}_{ij}=\sigma^{\prime}_{i}\sigma^{\prime}_{j}$, 
between measurements) are taken to contribute to the off-diagonal elements in $V^{\prime}$.
All error contributions (statistical, experimental, hadronization,
scale uncertainty and the uncertainty from massless calculations) are taken to 
contribute to the diagonal elements. 
The hadronization and scale uncertainties are computed
from the \as\ values obtained with the alternative hadronization models
and with the upper and lower values of the renormalization scale,
respectively, and combined according to Eq.~\ref{weight}.
The uncertainty for the massless calculations is also
combined according to Eq.~\ref{weight}.
\section{Systematic Uncertainties}
\label{systematic}
Several sources of possible systematic uncertainty are studied
as in~\cite{pr404}. All
systematic uncertainties are taken as symmetric. \par
{\bf Experimental uncertainties:}
Contributions to the experimental uncertainties are estimated by 
repeating the analysis with varied cuts or procedures.
For each systematic variation the value of \as\ is determined and
then compared to the result of the standard analysis (default value). 
In each case, the difference 
with respect to the default value is taken as a systematic uncertainty.
\begin{itemize}
\item[(1)]
The algorithm to avoid the double counting of energy in tracks and calorimetry is not
applied. Instead all observables are computed using all charged tracks 
and electromagnetic calorimeter clusters. The effects of double counting 
are then taken into account exclusively through the detector correction.
\item[(2)]
The containment cut is tightened to be well within the barrel, $|\cos(\theta_{\mathrm T})| < 0.7$.
\item[(3)]
Instead of using \py\ for the correction of detector effects as
described in Section~\ref{detectorcorrection}, events generated with \hw\ are 
used.
\item[(4)]
The fit range\footnote{The determination 
of the fit range is explained in section~\ref{fitprocedure}} is changed. 
Two different cases are considered. First  the fit range 
is decreased by one data point at each edge of the fit range. 
Second the fit range is extended by one data point at each
edge of the fit range. The larger deviation from the default fit is 
taken as a systematic uncertainty. 
In order to take statistical fluctuations into account, 
the deviation is calculated using the average deviation of a 
fit applied to 50 Monte Carlo samples.
\item[(5)]
The algorithm to compute $s'$ is replaced by a simpler version 
accounting for at most one initial-state photon~\cite{sellep2}.
\item[(6)]
The cut on the likelihood variable ${\cal L}_{\qqbarqqbar}$ to reject background
from $\ww \to \qqbarqqbar$ events is changed from  0.25 
to 0.1 and 0.4. The larger deviation from the default value is taken as the
systematic uncertainty.
\item[(7)]
The cut on the likelihood variable ${\cal L}_{\qqbar\lnu}$ to reject background
from $\ww \to \qqbar\lnu$ events is changed from 0.5 to 0.25 and 0.75. 
The larger deviation from the default value is taken as the
systematic uncertainty.
\item[(8)]
The amount of subtracted four-fermion background is varied by $\pm 5 \%$. The larger
deviation from the default value is taken as the systematic uncertainty.
\end{itemize}
Variation (5) applies only to the data taken above the \zzero\ resonance and variations
(6)-(8) only to data taken at and above the W-pair threshold at 161~GeV. 
All experimental uncertainties are added in quadrature and the result is quoted as the
experimental systematic uncertainty. 
None of the variations contribute dominantly to the overall 
experimental systematic uncertainty. \par
{\bf Hadronization:} 
The uncertainties associated with the hadronization correction 
are evaluated by using \hw\ and 
\ar\ instead of \py. The larger change in \as\ resulting
from these alternatives is taken as the error. \par 
{\bf Scale uncertainties:}
The uncertainty, associated with missing higher order terms in the
theoretical prediction, is assessed by varying the renormalization scale factor \xmu. The
predictions of a complete QCD calculation would be independent of \xmu, but a
finite-order calculation such as that used here retains some dependence on \xmu.
The renormalization scale \xmu\ is set to half and twice the value
of \xmu\ from the default fit. The larger deviation from the default 
value of \as\ is taken as the systematic uncertainty. 
The uncertainty band method as discussed in~\cite{uncband} cannot 
be applied to this analysis, because the rescaling of the 
NLLA terms and the use of different matching schemes
are currently not possible. However, the renormalization scale variation 
in the range $0.5 < \xmu < 2.0$ gives scale uncertainties for most
observables consistent with uncertainties obtained 
with the uncertainty band method~\cite{uncband}.\par
{\bf Uncertainties from massless QCD calculations:}
In the case of the four-jet rate an additional
uncertainty from massless QCD calculations is considered.
The QCD prediction used is only available
for the massless case. Effects from massive b quarks are
expected and the corresponding uncertainties on the
strong coupling are evaluated for the four-jet rate.
The overall four-jet rate can be written as the sum of the 
the four-jet rate originating from b quark and from light 
quark events, $R_{4}=f_{\mathrm{b}} R_{4}^{\mathrm{b}}+ (1-f_{\mathrm{b}})R_{4}^{\mathrm{light}}
=(f_{\mathrm{b}} B_{4} + (1-f_{\mathrm{b}}))R_{4}^{\mathrm{light}}$,
with $f_{\mathrm{b}}$ being the fraction of b quark events, 
$R_{4}^{\mathrm{b}}$ the four-jet rate originating from
b quark events, $R_{4}^{\mathrm{light}}$ the four-jet rate 
originating from light quark events and 
$B_{4}= R_{4}^{\mathrm{b}}/ R_{4}^{\mathrm{light}}$. 
The Standard Model expectation for the fraction $f_{\mathrm{b}}$ computed 
in ~\cite{zfitter} varies between $0.216$ for data taken at 
LEP1 and $0.169$ at 207 GeV. Following the studies of QCD heavy 
mass effects~\cite{bmass} $B_{4} \approx 0.9$ at $\rs=\mz$, leading to 
a change in the overall four-jet rate $R_{4}$ of $\approx 2\%$.
Even though the effect is expected to decrease with increasing \rs,
a value of $B_{4}=0.9$ is used as a conservative estimate 
for the whole energy range. Since $R_{4}\sim \assq $ in LO QCD we 
have $\Delta \as / \as =  \Delta R_{4} /2 R_{4}$ and thus 
we expect a $\approx 1\%$ relative change in \as. 
This is taken to be our massless QCD uncertainty.
\section{Results}
\subsection{Four-Jet Rate Distributions}
\begin{figure}[ht]
\centerline{
\epsfxsize=8cm\epsfbox{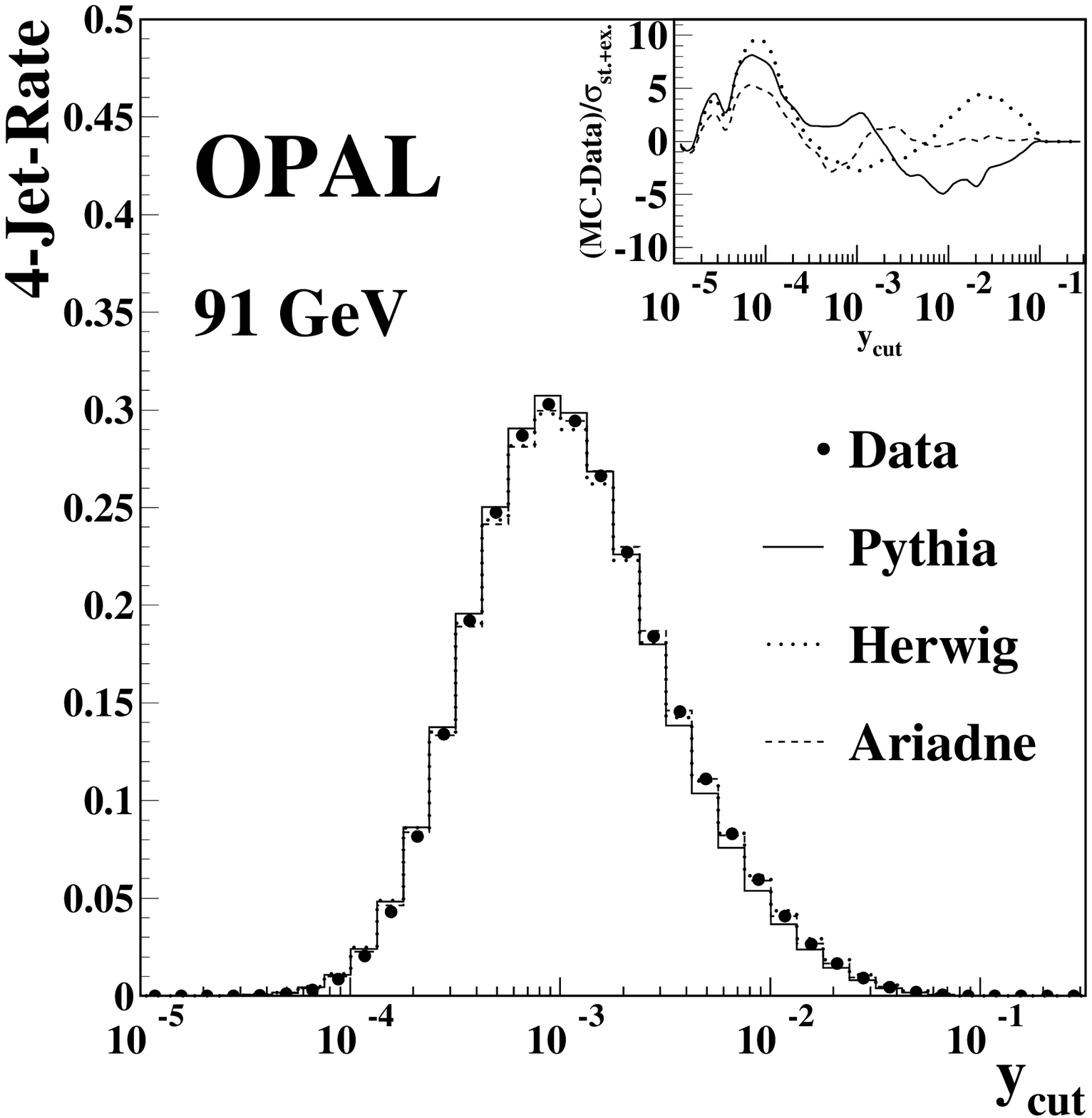}
\epsfxsize=8cm\epsfbox{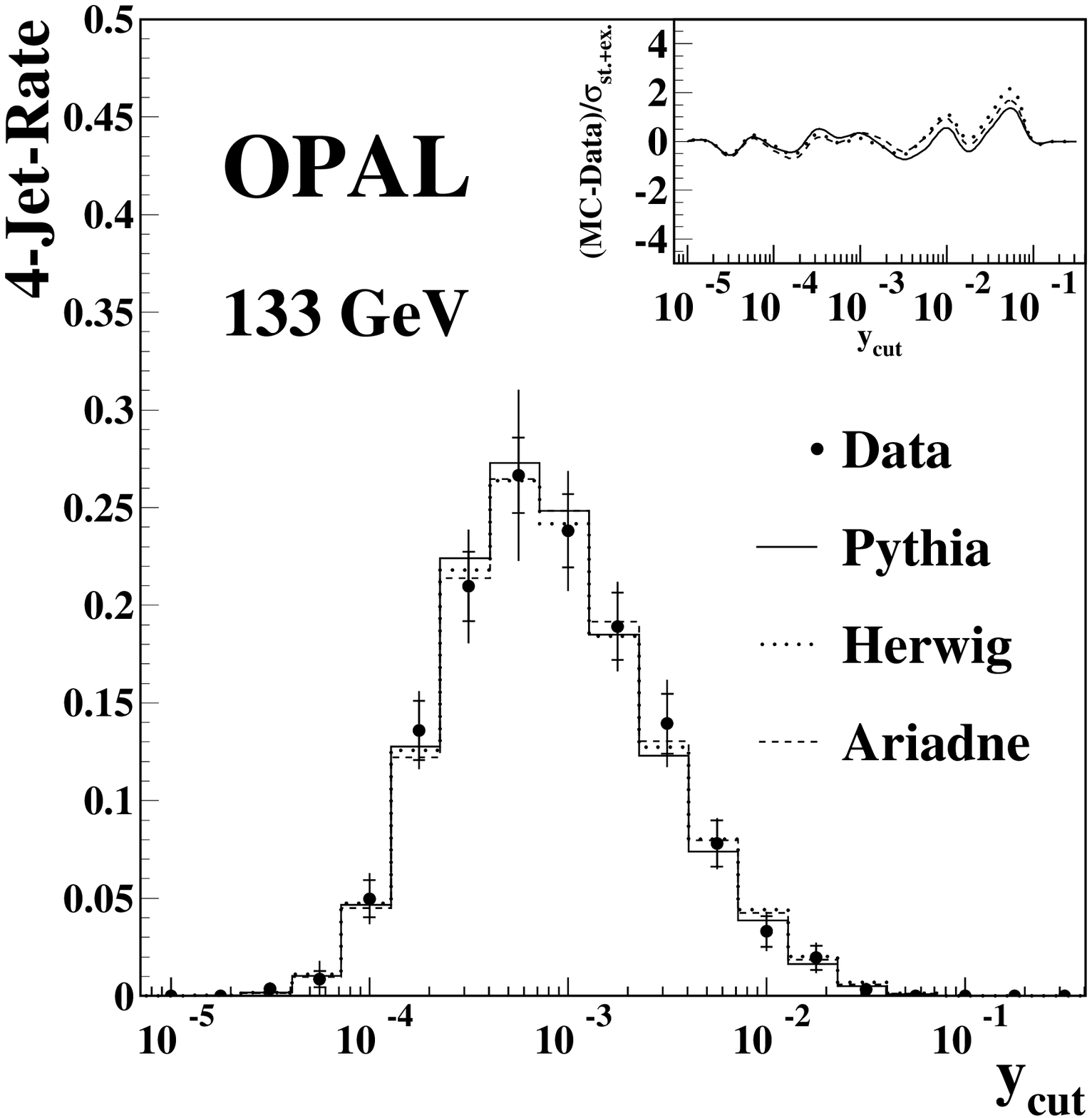}}
\centerline{
\epsfxsize=8cm\epsfbox{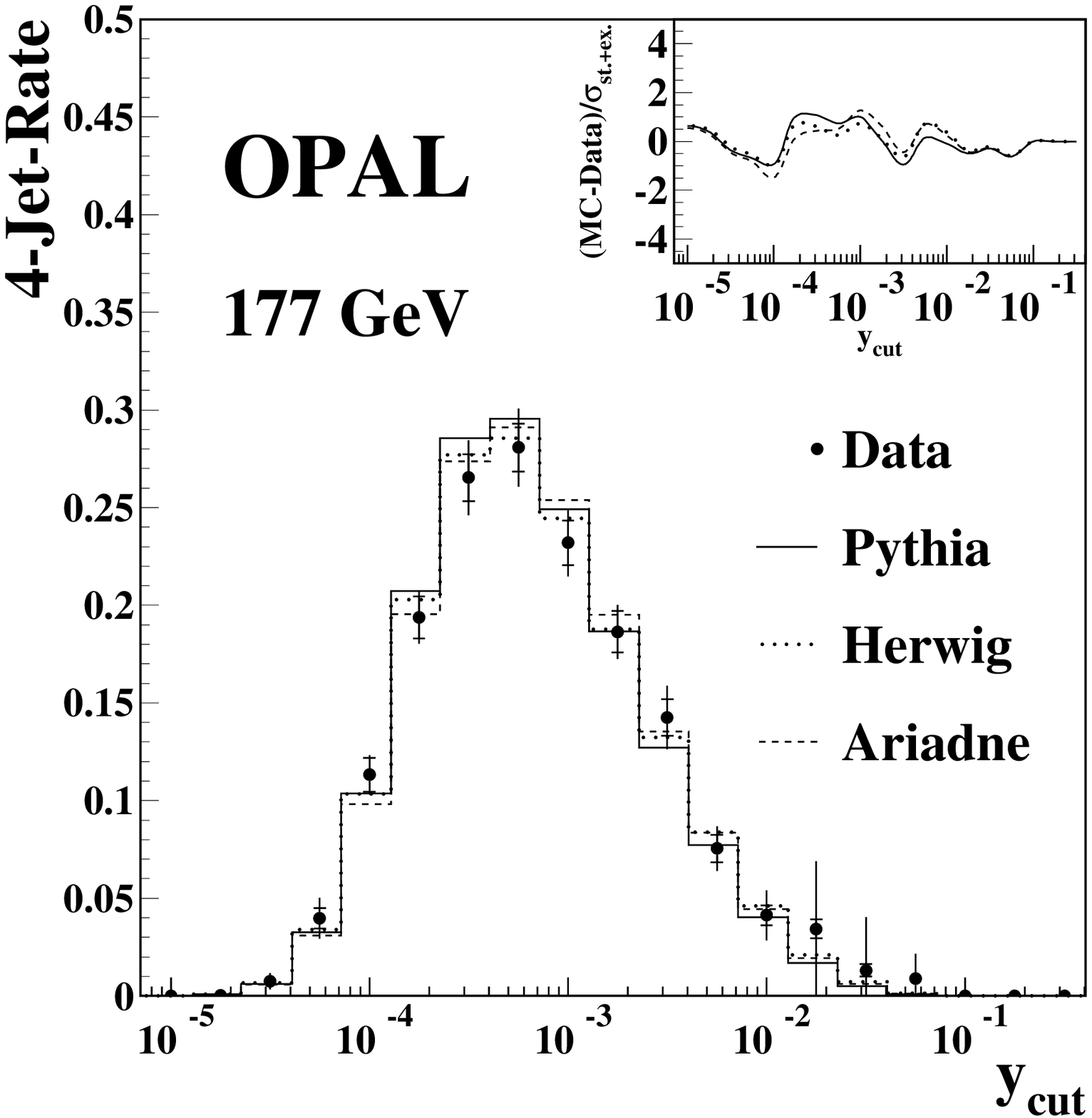}
\epsfxsize=8cm\epsfbox{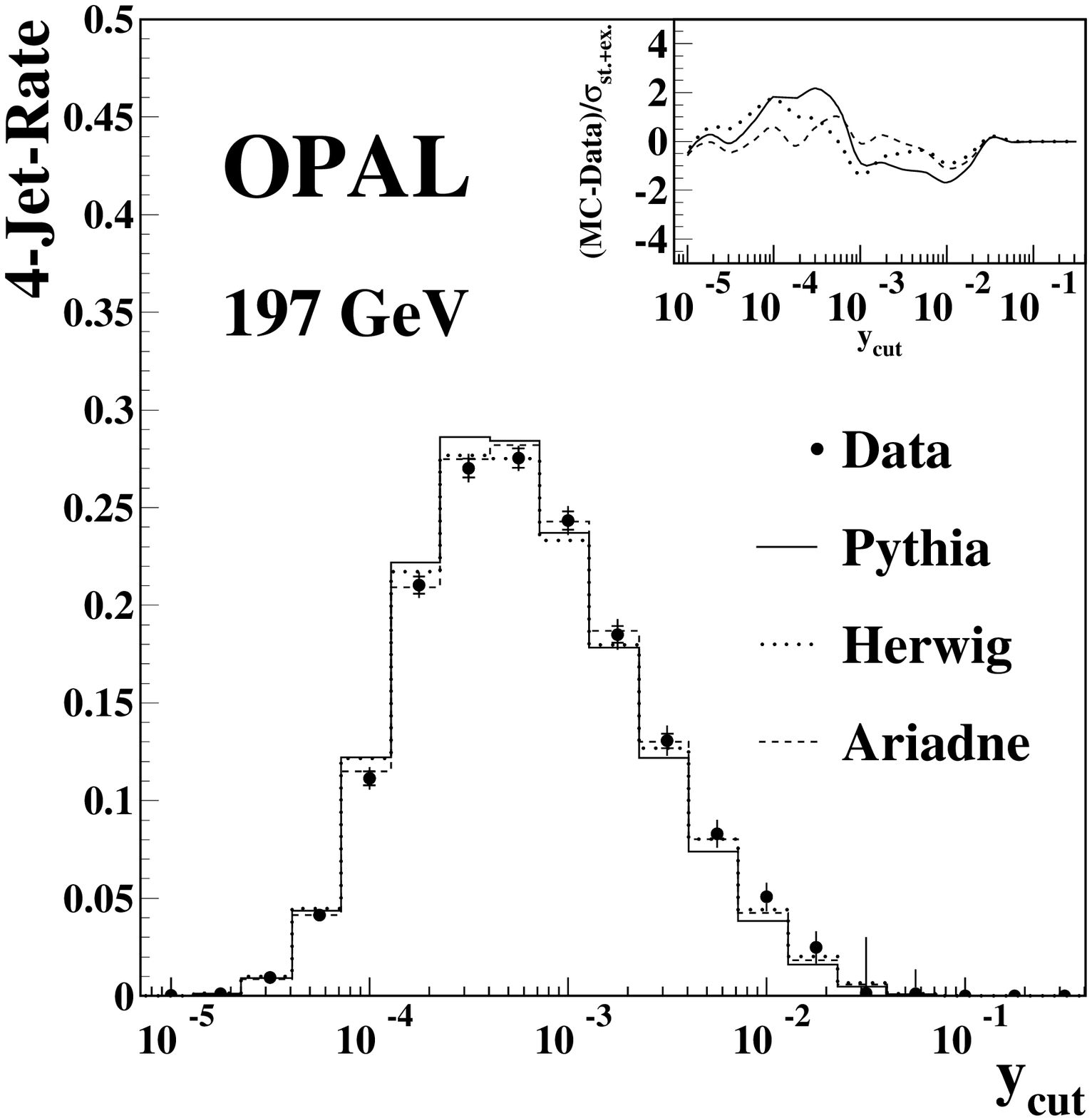}}
\caption{The four-jet rate distribution
at hadron level as a function of the $y_{\mathrm{cut}}$ resolution
parameter obtained with the Durham algorithm. 
The four-jet rates are shown for the data corrected to the hadron level
at four average centre-of-mass energies 
between 91 and 209~GeV together with predictions based on \py, \hw\ and \ar\ Monte 
Carlo events generated  at the averaged energy. 
The error bars show the statistical (inner error bars) and 
experimental uncertainties added in quadrature.  
Error bars not shown  are smaller than the point size. 
The panel in each upper right corner shows the differences between 
data and Monte Carlo predictions, divided by the sum of the statistical 
and experimental error. For data points with no data events, the difference
is set to zero. }
\label{hadron}
\end{figure}

The four-jet rates for the four energy intervals, after subtraction 
of background and correction for detector effects, are shown 
in Figure~\ref{hadron}. 
Superimposed we show the distribution predicted by the \py, \hw\ and \ar\ Monte
Carlo models.
In order to  allow for a more quantitative comparison
between data and models, the inserts in the upper right corner show 
the differences between data and each model, divided by the 
combined statistical and experimental error 
in each data point The sum of
squares of these differences
would, in the absence of correlations,
represent a $\chi^{2}$ between
data and the model. However, since correlations are present, such $\chi^{2}$ 
values should be regarded only as a rough indication of the agreement between
data and the models. The three models are seen to describe the high
energy data well. Some discrepancies are, however, seen in the 91~GeV data
which have a much better precision. The different Monte Carlo
models do not agree with each other; \ar\ tends to give the best 
description of the data. 
\subsubsection{Determination of \boldmath{\as}}
\label{fitprocedure}
\begin{figure}[ht]
\centerline{
\epsfxsize=8cm\epsfbox{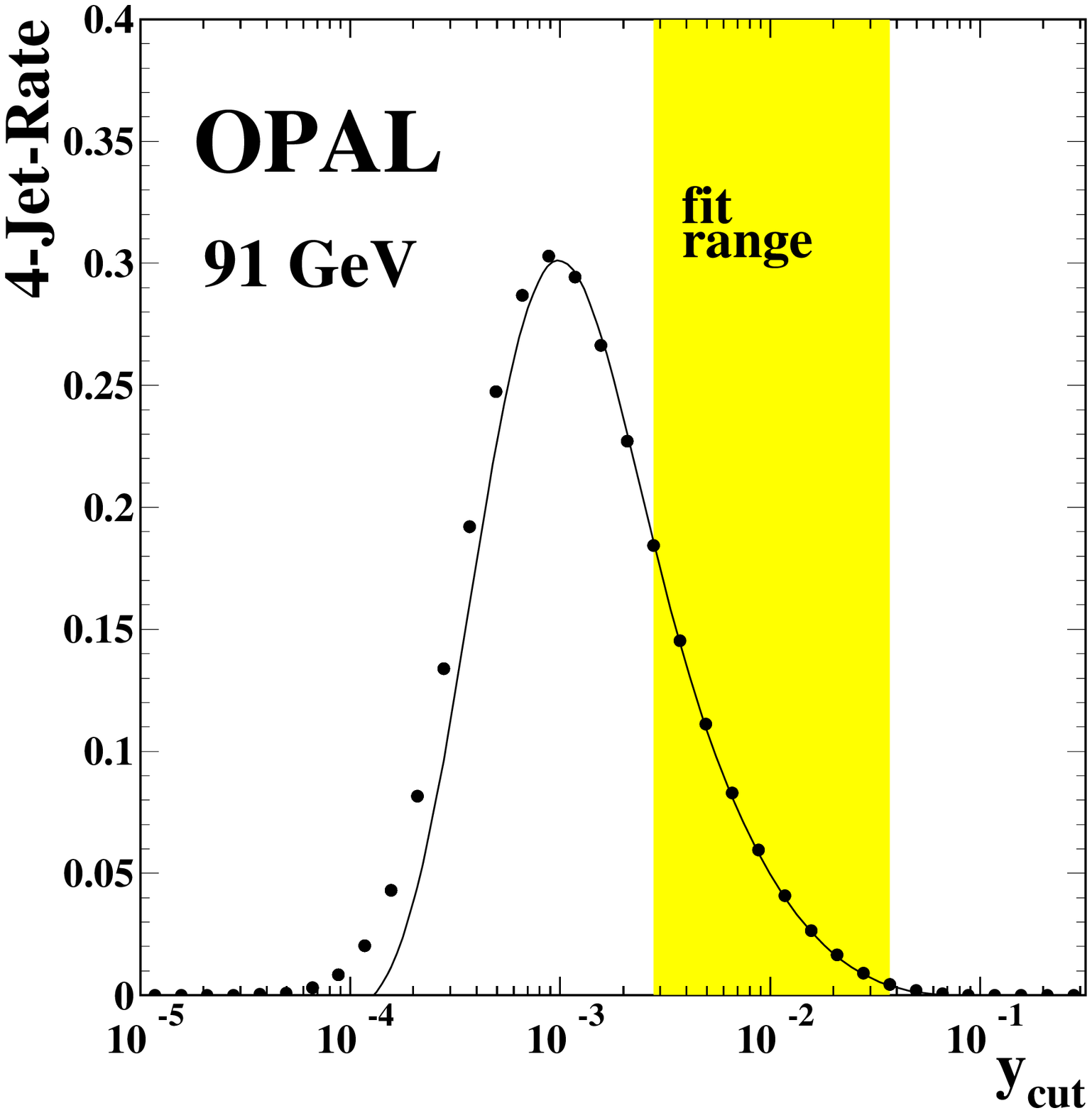}
\epsfxsize=8cm\epsfbox{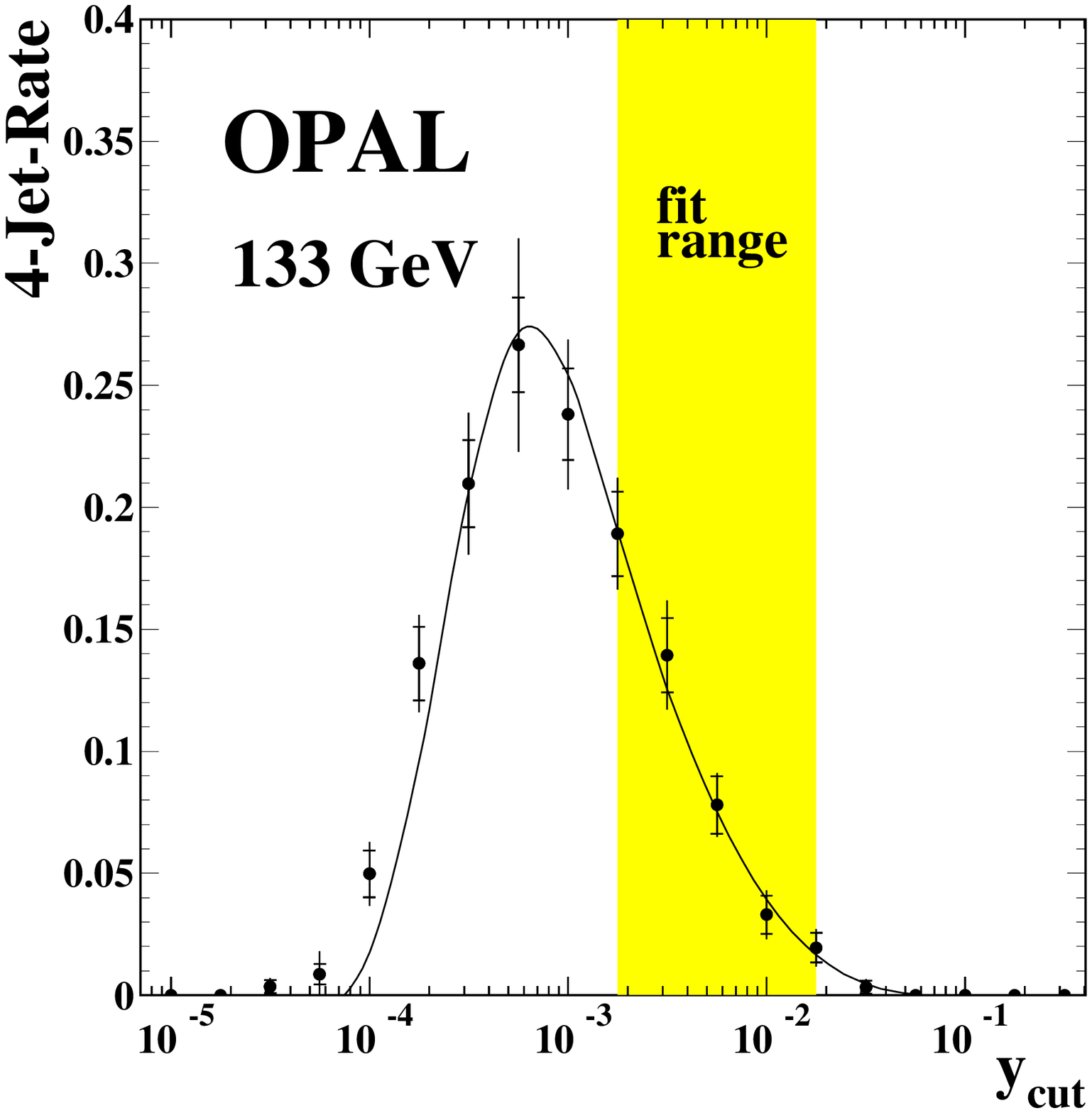}}
\centerline{
\epsfxsize=8cm\epsfbox{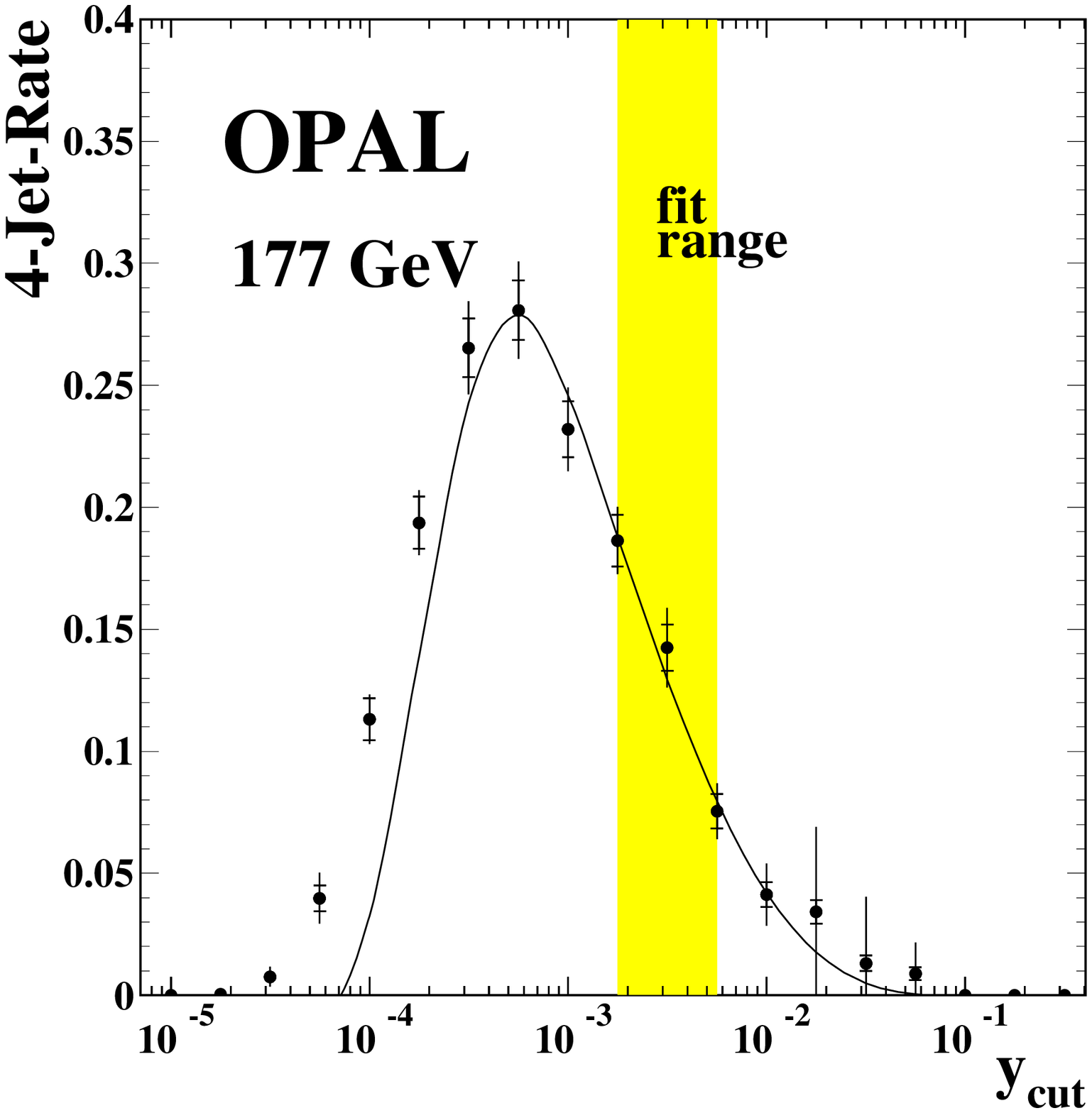}
\epsfxsize=8cm\epsfbox{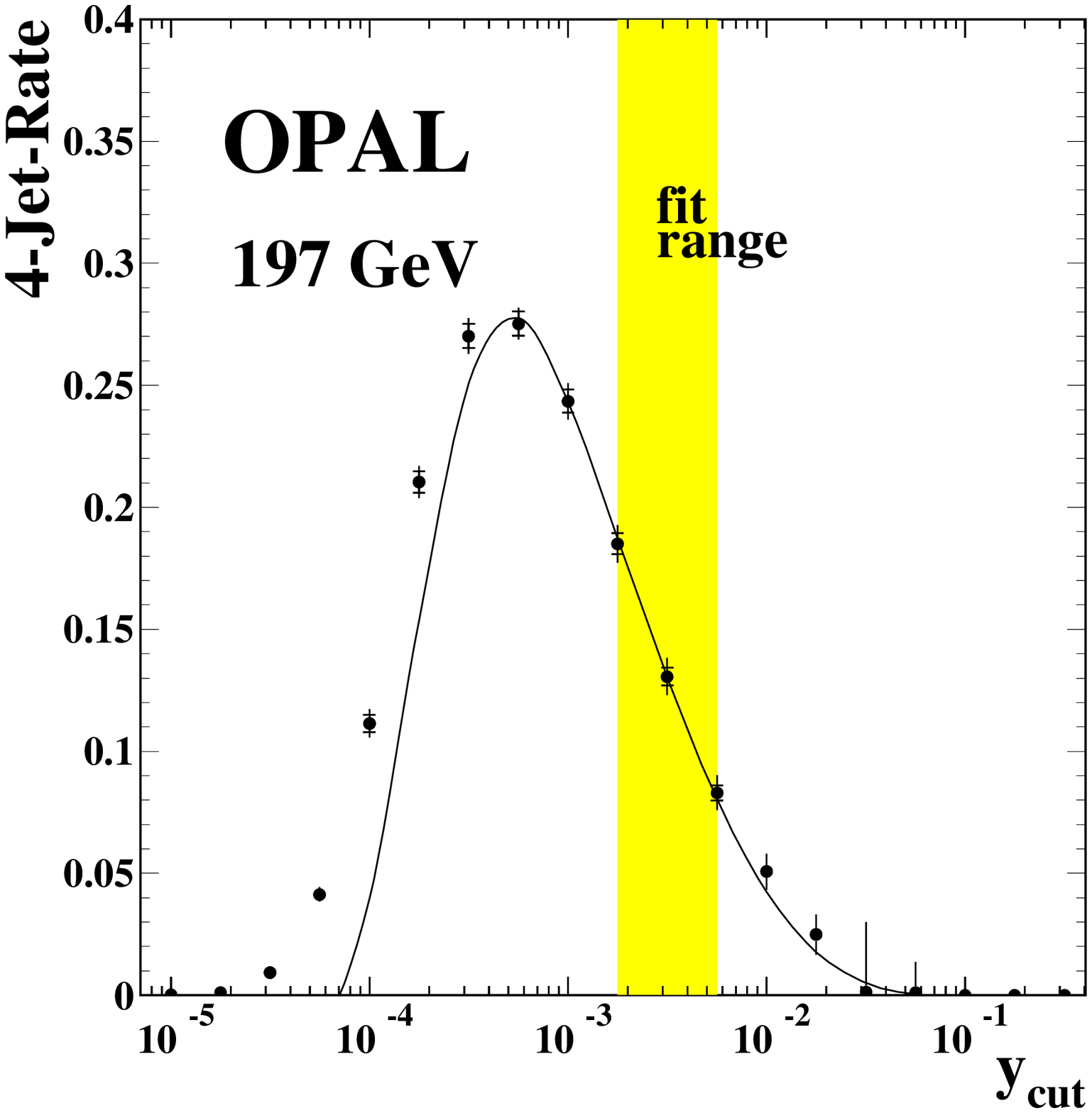}}
\caption{
The hadron-level four-jet rate distributions 
for energies between 91~GeV and 209~GeV. 
The error bars show the statistical (inner error bars) and
experimental uncertainties added in quadrature.
When not shown, the errors are smaller than the point size.
The curves show the theory prediction after $\chi^{2}$ minimization 
within the fit range indicated. The fit range is determined 
as discussed in Section~\ref{fitprocedure}. The data points
are strongly correlated and an enlarged fit range would 
lead to only a minor gain in statistical precision.}
\label{fit_plot}
\end{figure}

\begin{figure}[ht]
    \begin{center}
    {\includegraphics[width=1.0\textwidth]{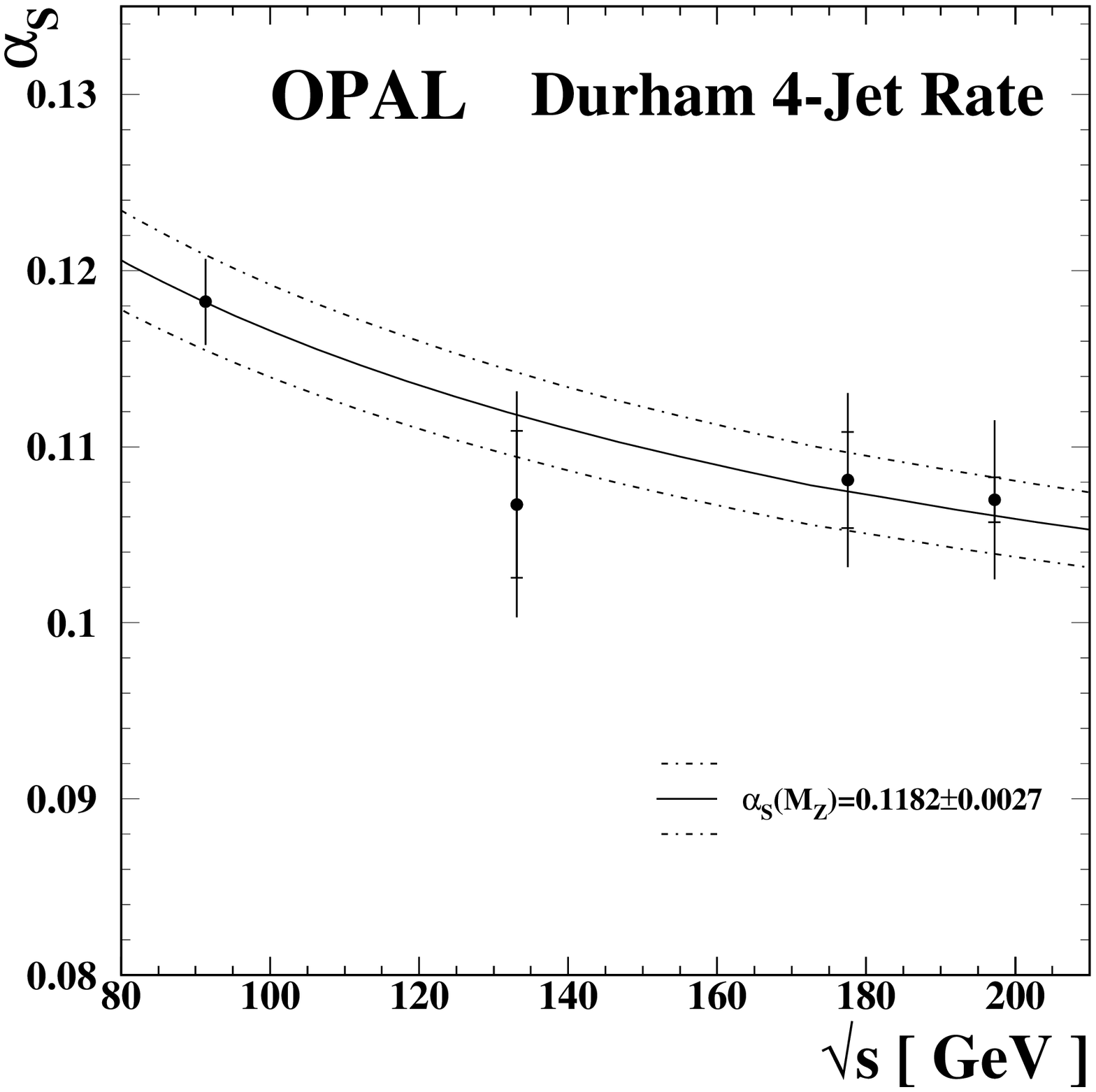}}
\end{center}
\caption{The values for \as\ obtained by a fit to the four-jet rate
with \xmu\ set to 1.0 in the four energy intervals.
The error bars
show the statistical (inner error bars) and the total error. The statistical
error at 91~GeV is smaller than the point size.
The lines indicate the current world average from~\cite{alphasbethke}
with the one standard deviation uncertainty.}
\label{alphas_fit}
\end{figure}

For the QCD prediction of $R_{4}(\ycut)$ the combined $\cal{O}(\ascu)$+NLLA 
calculation (Eq.~\ref{NLLA}) is used. 
The scale parameter $\xmu$ is set to one. 
The fit range is  $0.0028 < \ycut < 0.0371$ for data taken at 91~GeV,
$0.0018 <  \ycut < 0.0178$ for data taken at LEP1.5  and 
$0.0018 <  \ycut < 0.0056$ for data taken at LEP2. 
The fit range is determined by requiring that the hadronization 
corrections be less than $20\%$ and the
detector corrections be less than $50\%$.
As shown in~\cite{zoltan} the theoretical 
uncertainty due to higher order missing terms
increases significantly for \ycut\ values below $0.005$. 
Therefore we impose  a further constraint on the \ycut\ region -- that 
the theoretical  prediction should not vary by more than $5\%$ as 
the renormalization scale factor \xmu\ is varied from 0.5 to 2~\footnote{
The uncertainty of the hadronization models and 
the detector correction using different Monte Carlo models is 
also less than $5\%$.}.
The fit range for the LEP2 data contains only three data points;
an enlarged fit range would lead to only a minor gain in
statistical precision.
We require the same fit range for all energy points
above 136~GeV.
In Figure~\ref{fit_plot} the hadron-level four-jet rate 
distributions for the four energy intervals are shown together with 
the fit result. The fit ranges cover the falling slope 
of the distributions at large \ycut, where the perturbative QCD
predictions adequately describe the data.
The increasing slope of the distributions at low \ycut\ values 
is less well described by the perturbative QCD description.
In this region 
the hadronization corrections become large (more than $100\%$) 
and the various hadronization models start to differ significantly.
The fits performed at individual
energy points are combined in the four different energy 
bins: 91.2~GeV; 130 and 136~GeV; 161, 172 and 183~GeV; and 189--209~GeV 
using the method described in Section~\ref{combination}.
The results are summarized in Table~\ref{fitcombined}
and shown in Figure~\ref{alphas_fit}.  
At LEP1 the statistical uncertainty is much smaller than
the theoretical one. 
The hadronization scales  as an inverse power of $\sqrt{s}$ and 
therefore the hadronization uncertainty decreases with
increasing energy. 
The fit range for data taken at LEP2 extends towards lower \ycut\ 
values compared to the fit range for data taken at LEP1. This
leads to an increased theoretical uncertainty for \as.
\begin{table}[h]
\begin{center}
\begin{tabular}{|r|r|r|r|r|r|r|r|r|} \hline
average $\sqrt{s}$ & \as & stat. & exp. & hadr. & scale & mass\\ 
in GeV & & & & & & \\
\hline
 $91$  & $ 0.1182 $ & $ 0.0002 $  & $ 0.0013$ & $ 0.0012$ & $ 0.0011$ & $0.0013$  \\
 $133$ & $ 0.1067 $ & $ 0.0042 $  & $ 0.0045$ & $ 0.0010$ & $ 0.0011$ & $0.0010$ \\
 $177$ & $ 0.1081 $ & $ 0.0027 $  & $ 0.0037$ & $ 0.0007$ & $ 0.0013$ & $0.0009$ \\
 $197$ & $ 0.1070 $ & $ 0.0013 $  & $ 0.0039$ & $ 0.0006$ & $ 0.0016$ & $0.0009$ \\
\hline
\end{tabular}
\end{center}
\caption{The mean value of \as\ for each energy interval, the statistical and
experimental errors, and the errors due to hadronization and 
scale uncertainties and massless QCD. }
\label{fitcombined}
\end{table}
Finally we combine all measurements to a single value of \asmz, with the final
result:
\be
\result, 
\ee
which is clearly dominated by the \as\ value
obtained at 91~GeV. The weight as calculated 
in Eq.~\ref{weight} for the data taken at 91 GeV is $84.3\%$.
\subsubsection{\boldmath Scale Dependence of \as }
\label{r4scalefree}
For the fits in Section~\ref{fitprocedure} the renormalization 
scale is set to the natural choice $\xmu=1$. However, different schemes for 
the determination of the renormalization scale are proposed~\cite{xmuscheme}. 
In this section we investigate two of these.\par
\begin{figure}[ht]
    \begin{center}
    \includegraphics[width=1.\textwidth]{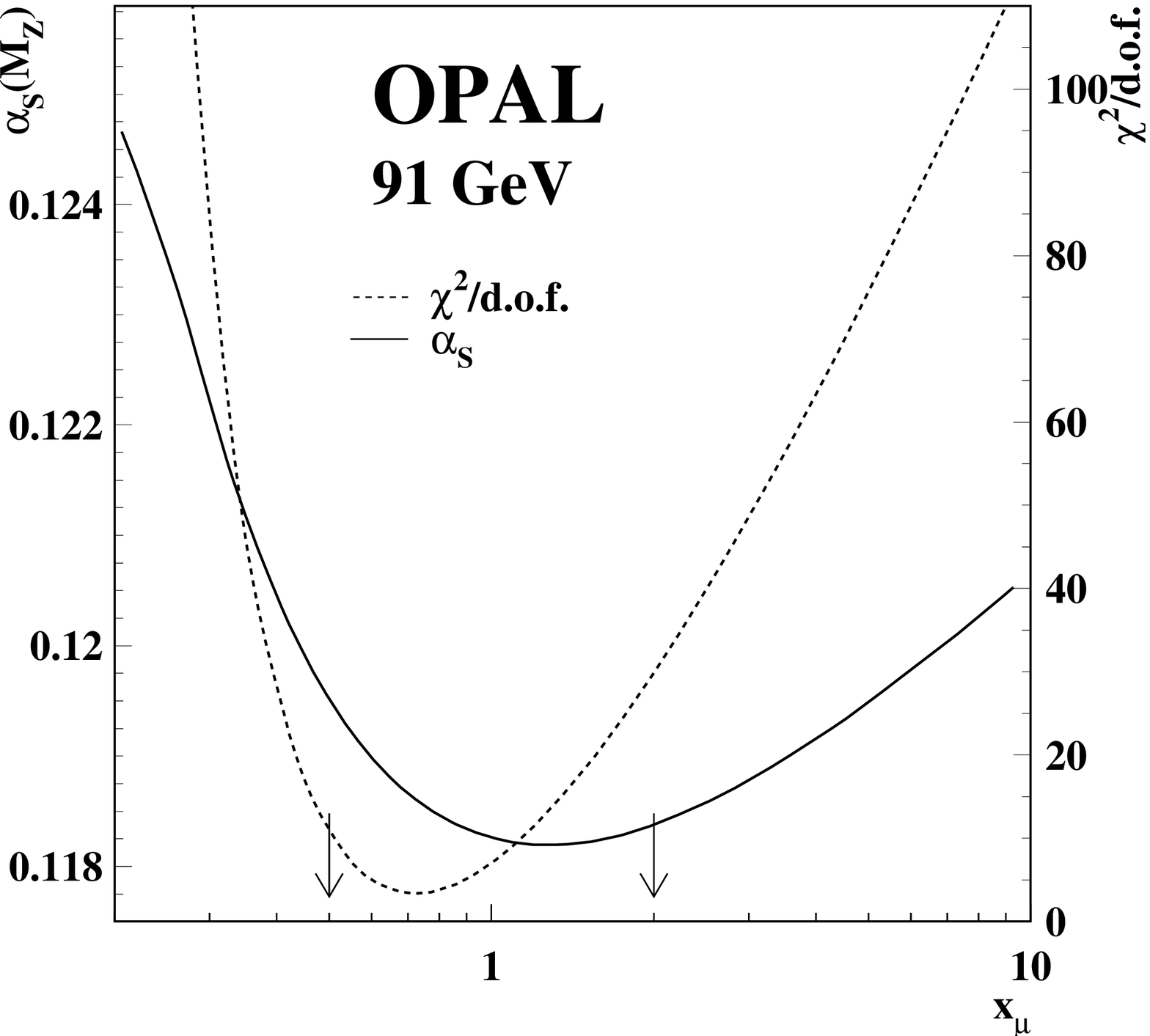}
\end{center}
\caption{The values of \as\ and \chisqd\ from the fit to 
the four-jet rate at 91 GeV data 
as a function of the scale parameter \xmu. 
The arrows indicate the variation $0.5 < \xmu < 2.0$
used to determine the theoretical systematic uncertainty.}
\label{xmuopt}
\end{figure}

In the optimized renormalization scheme, discussed in detail 
in~\cite{xmuopt}, the minimization is performed 
with both \as\ and \xmu\ treated as free parameters. 
However, at some low-statistics LEP2 energy points the fits 
did not converge. For this reason we applied a
modified version of the optimized renormalization scheme. 
The optimized scale \xmuopt\ was determined with 
the high statistics sample at 91~GeV and then the same scale was
used at higher energies.
The result for \xmuopt\ at 91~GeV is \rxmufree. 
The variation of \chisqd\ as a function of the scale \xmu\ for 
91~GeV data is shown in Figure~\ref{xmuopt}.
The combination of all energy points using the method described 
in Section~\ref{combination} returns a value of
\be
\resultxmufree.
\ee
The weight as calculated in Eq.~\ref{weight} for the data taken
at 91 GeV is $73.1\%$, leading to an increased statistical
uncertainty in the combined result. \par
The second choice for the determination of the renormalization scale 
followed approximately the approach suggested by Stevenson~\cite{scaleasmin}. 
The renormalization
scale \xmumin\ is specified by the point with \as\ having the least sensitivity 
to the renormalization scale \xmu. 
Using the 91 GeV data (Figure~\ref{xmuopt}) \xmumin\ is determined to be \rxmumin.
The fit is repeated at all other energy points with \xmu\ set 
to \xmumin. The combination of all energy points using the 
method described  in Section~\ref{combination} returns a value of
\be
\resultxmuasmin.
\ee
Both alternative choices for the renormalization scale return a 
value of \as\ which
is well within the variation of the systematic uncertainty due to
missing higher order terms.
\subsection{Thrust minor and \boldmath{$D$-parameter}}
\label{eventshapes}
For the \tmi\ and $D$-parameter event-shape variables 
only the basic NLO calculations are available. For this reason and 
due to the problems discussed in the introduction we expect
large systematic uncertainties due to missing higher 
order terms. 
The strong coupling \as\ has not previously been determined
using these four-jet event-shape variables.
Here we perform a first measurement of \as\ to 
quantify the various experimental, hadronization and
theoretical uncertainties. \par
\begin{figure}[ht]
\centerline{
\epsfxsize=8cm\epsfbox{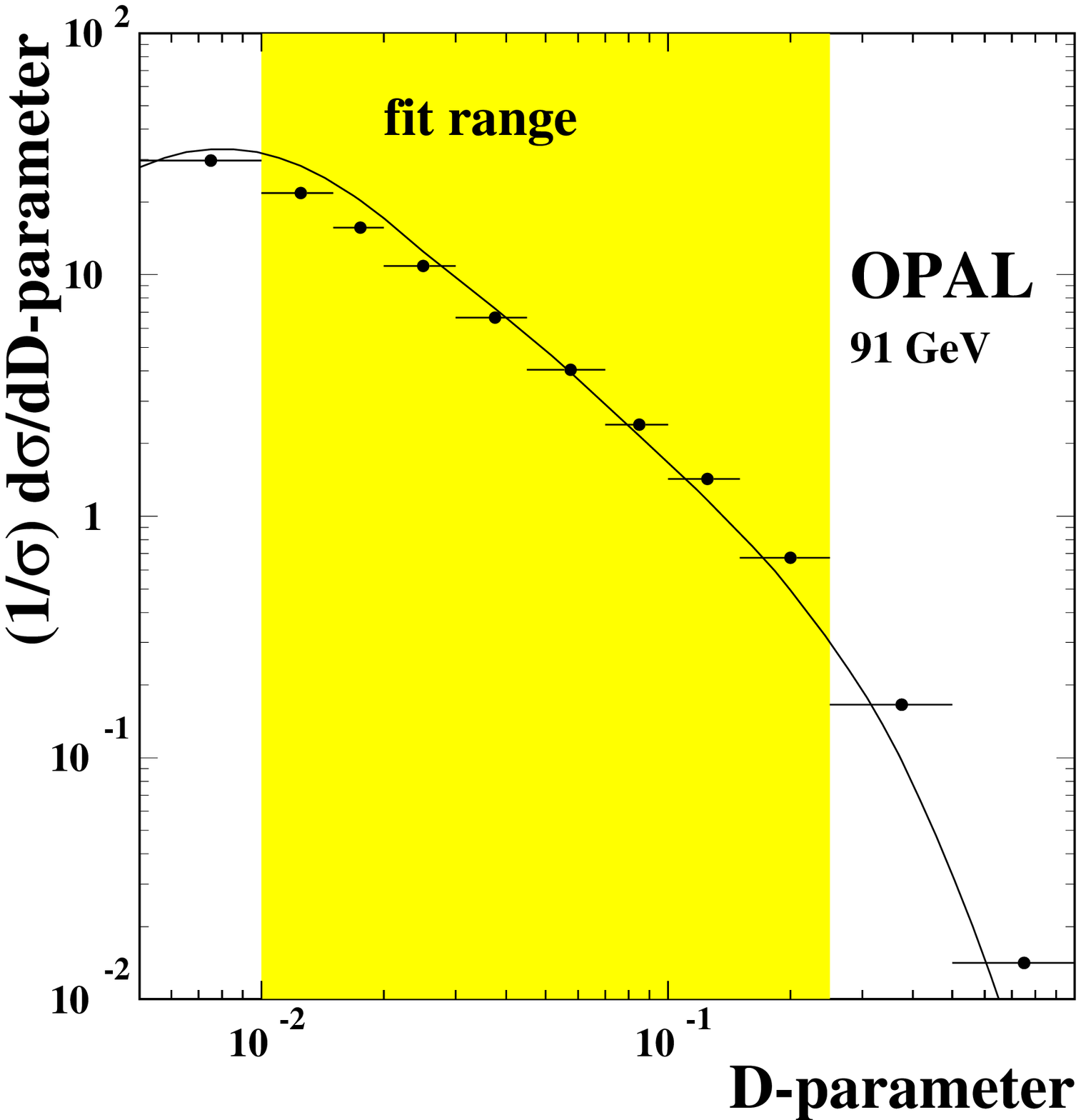}
\epsfxsize=8cm\epsfbox{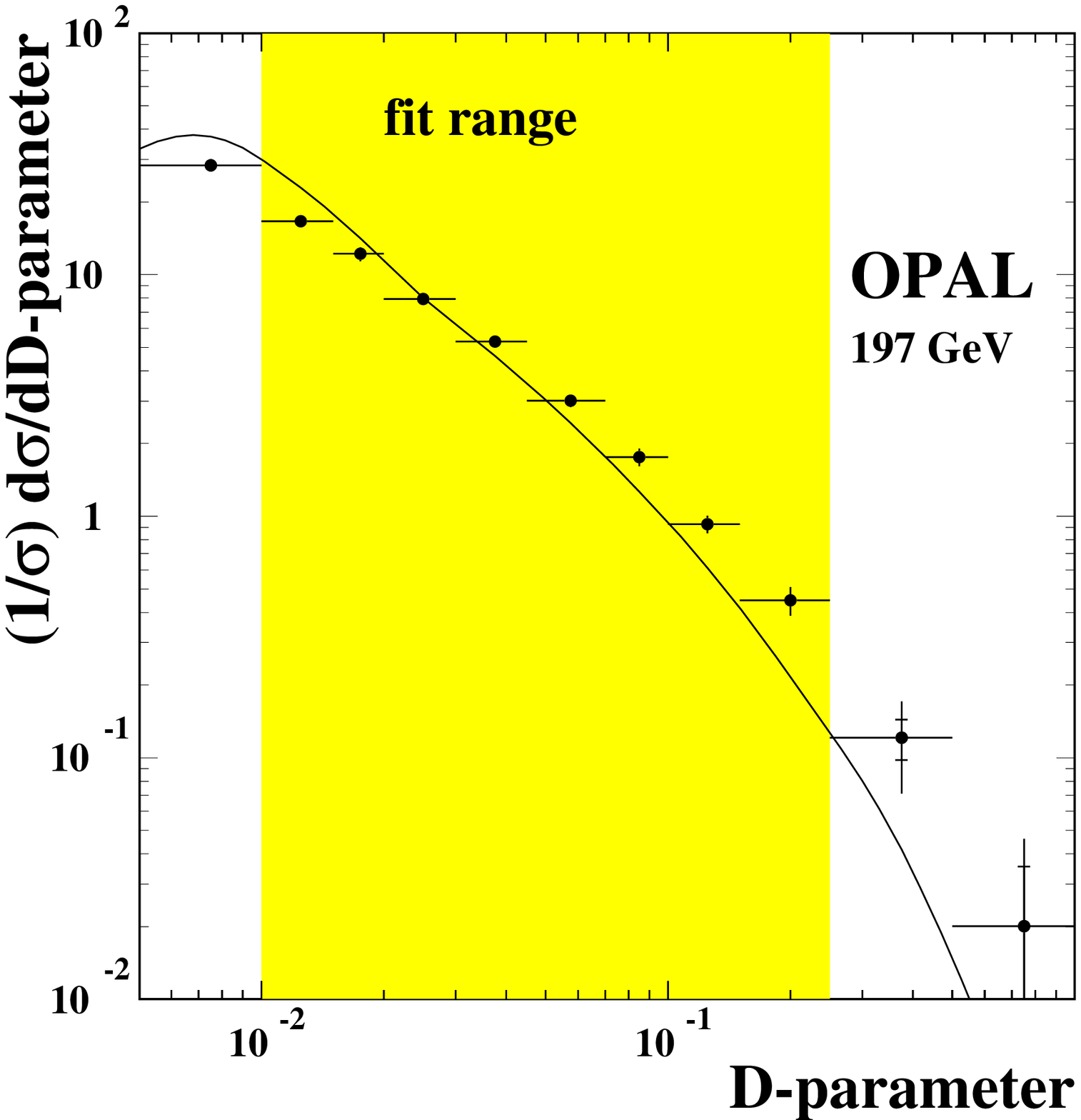}}
\centerline{
\epsfxsize=8cm\epsfbox{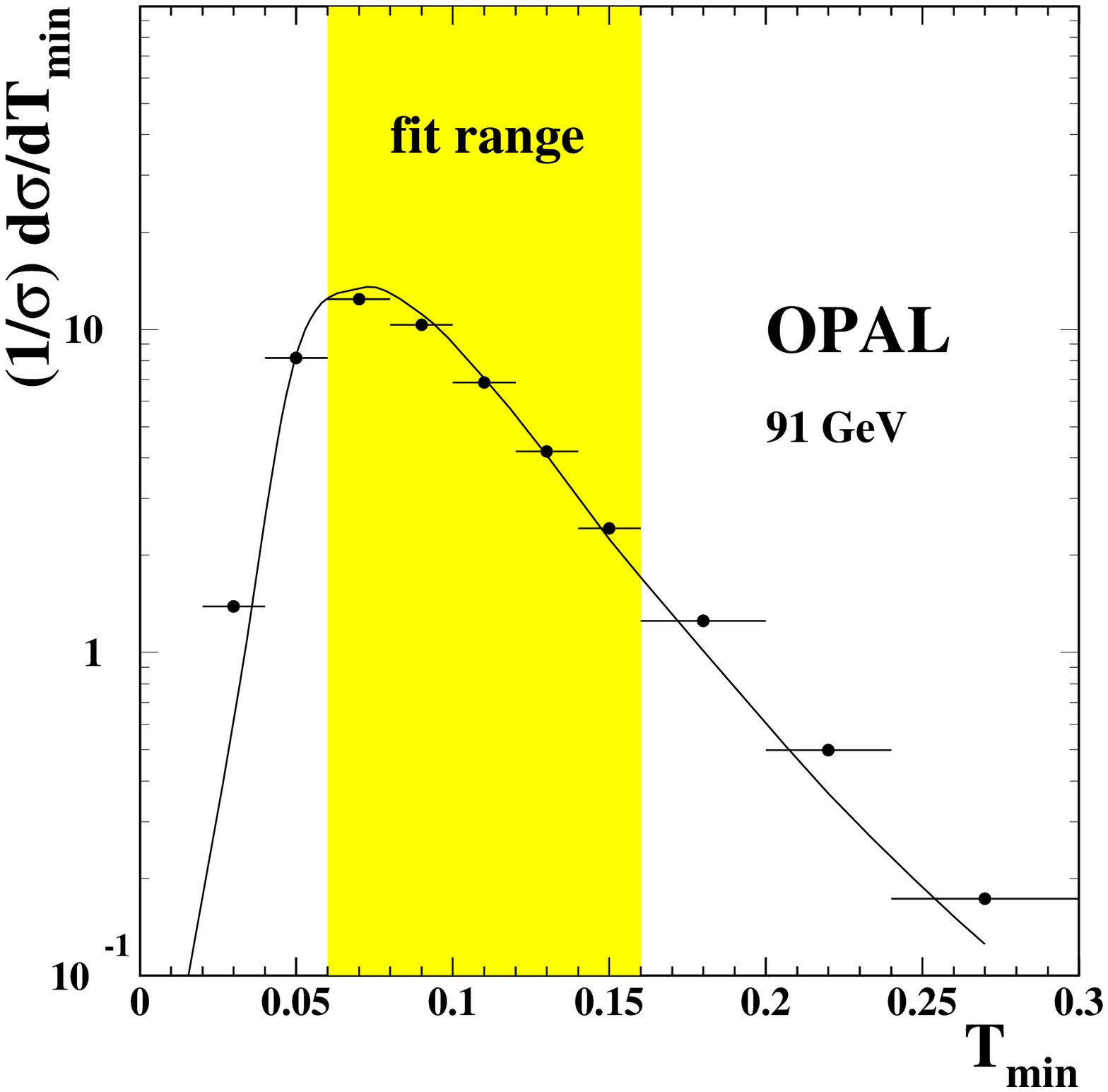}
\epsfxsize=8cm\epsfbox{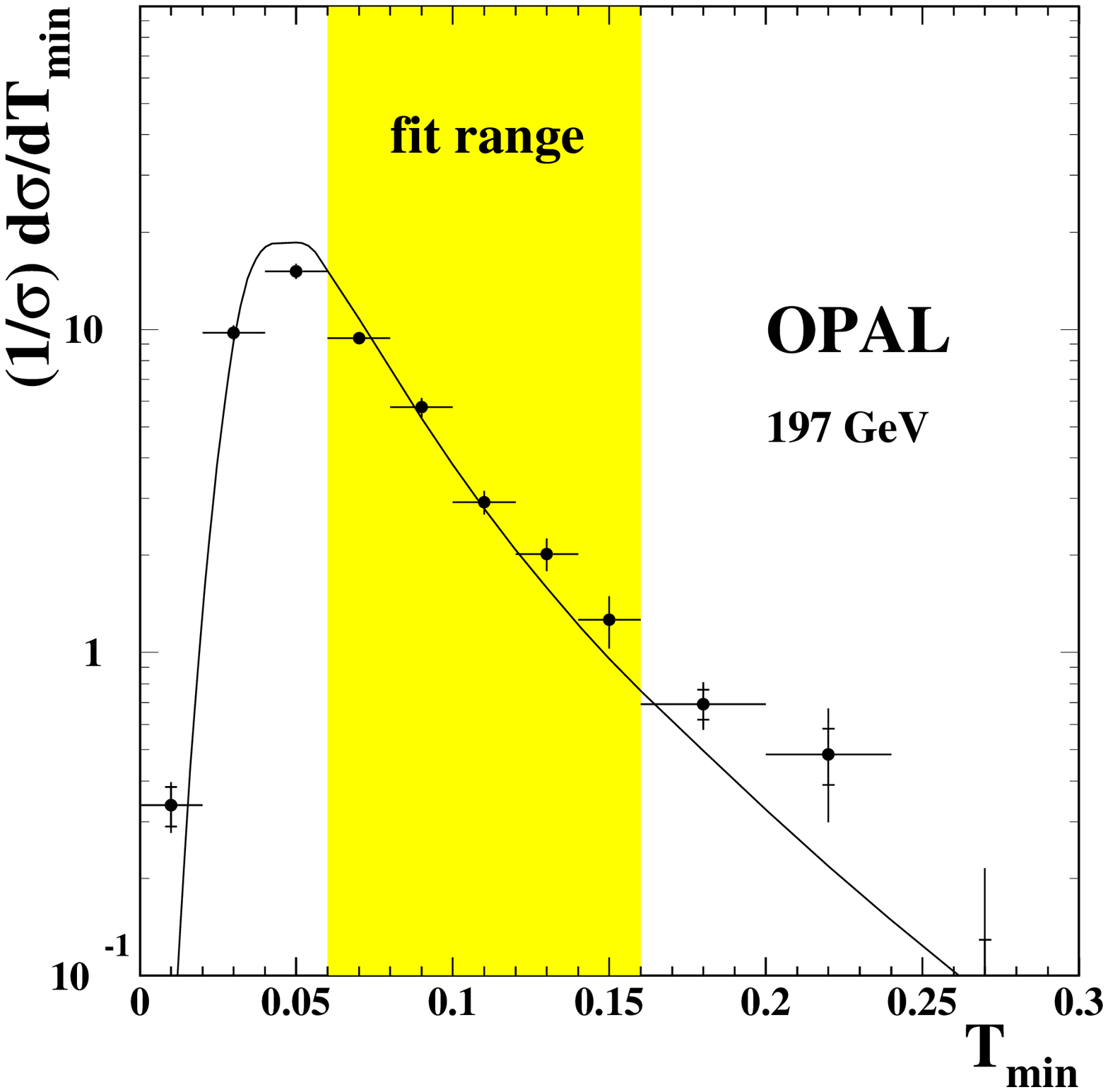}}
\caption{
The hadron-level $D$-parameter and \tmi\ distributions 
for energies of 91~GeV and 197~GeV. 
The error bars show the statistical (inner error bars) and
experimental uncertainties added in quadrature.
When not shown, the errors are smaller than the point size.
The curves indicate the theory prediction
after $\chi^{2}$-minimization with the renormalization scale set to 1.0.
The fit range is determined as discussed in Section~\ref{eventshapes}.}
\label{fit_dp_tm}
\end{figure}

Studies with the basic NLO calculations were already performed 
with three-jet event-shape observables~\cite{xmuopt,delphiscale}. 
These investigations showed that fits with the renormalization 
scale fixed to $\xmu=1$
described the data worse than fits with both 
parameters \as\ and \xmu\ free, which 
is usually interpreted to mean that
the optimized scale-choice mimics the effect
of higher order corrections. 
We use both approaches here.
The $D$-parameter and \tmi\ event-shape distributions were published 
in~\cite{pr404} and are used here.
For the QCD predictions we use the \debr\ $\cal{O}(\ascu)$ 
calculations~\cite{zoltan}.
In the first approach the scale parameter $\xmu$, as discussed in 
Section~\ref{theorydp}, is 
set to one.  
The fit range is set from 0.010 to 0.25 for the $D$-parameter and from 
0.06 to 0.16 for \tmi. 
The fit range is determined by requiring that the hadronization and the 
detector correction be less than $50\%$.
The results obtained at average centre-of-mass energies of 91~GeV and 197~GeV
are shown in Figure~\ref{fit_dp_tm}. 
For the $D$-parameter the slope of the curve obtained from the theory prediction 
after the $\chi^{2}$ minimization does not give a good description of the data.  
Similar effects were already seen previously with three- and 
four-jet observables ~\cite{campbell,delphir4,xmuopt,delphiscale}.
The results for the fit to the data between 91 and 209~GeV are combined to a single 
value for \asmz\ using the method introduced in Section~\ref{combination}. 
The result for the $D$-parameter
is 
\be 
\resdpxmu 
\ee 
and for \tmi
\be 
\restmxmu,
\ee 
both consistent with the result from the four-jet rate, albeit with large errors.
All uncertainties are greater than those on the four-jet rate determination. 
In particular the systematic uncertainty due to higher order missing
terms is significantly larger. \par
The impact of the scale \xmu\ on the result is studied with
a revised fit using the modified optimized renormalization scheme as in 
Section~\ref{r4scalefree}. 
The optimized scale \xmuopt\ is
determined with 91~GeV data only and the fit repeated with 
\as\ free and the scale set to \xmuopt. For the $D$-parameter
the scale is set to \dxmufree\ and for \tmi\ to \txmufree\footnote{
Other studies with observables using NLO predictions only with an optimized 
scale also result in \xmu\ values well below unity~\cite{xmuopt,delphir4}.}.
The description of the slope of the $D$-parameter improves significantly
when the optimized renormalization scale parameter is used.
The results are combined to a single value for \asmz\ using the method
introduced in Section~\ref{combination}. 
The result for the $D$-parameter is 
\be 
\resdp
\ee 
and for \tmi\ 
\be 
\restm.
\ee 
Due to the problems discussed above we do not 
combine the \as\ values based on the event-shape distributions with 
the one from the four-jet rate.  The differences between the result 
evaluated at the natural scale and at the optimized scale 
suggests, as discussed in~\cite{campbell,dparbad}, that the uncalculated higher order
terms are important. An improved
theoretical understanding is necessary before \as\ measurements 
using four-jet event-shapes are competitive.
\section{Summary}
In this paper we present the  measurements by the \Opal\ Collaboration 
of the strong coupling \as\ from the four-jet rate 
at centre-of-mass energies between 91 and 209~GeV.
The predictions of the \py, \hw\ and \ar\ Monte Carlo models are found
to be in general agreement with the measured distributions. \par
\begin{figure}[h]
  \begin{center}
    {\includegraphics[width=1.0\textwidth]{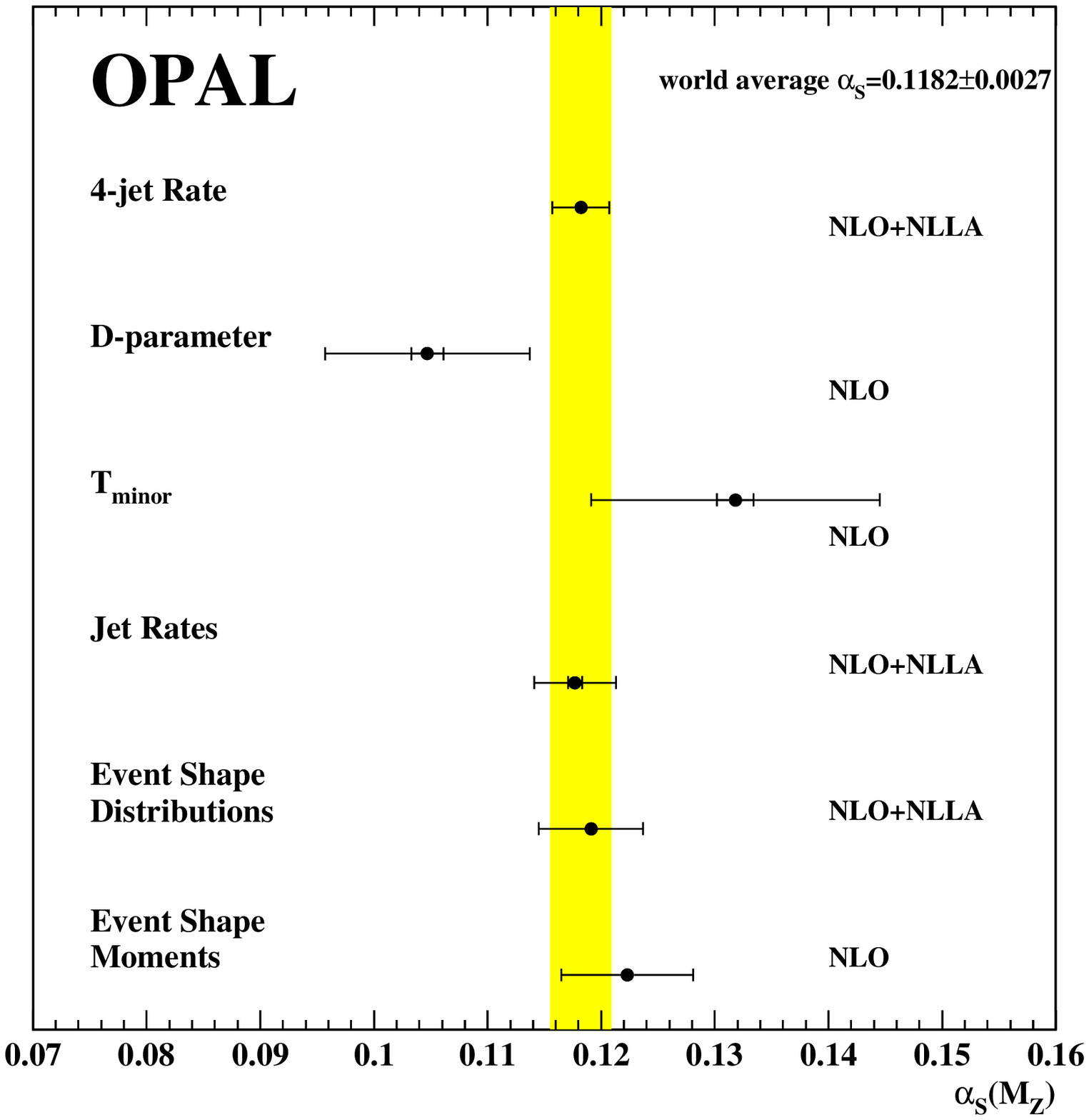}}
  \end{center}
  \caption{The results of the fit to the four-jet rate, the $D$-parameter and
    \tmi. The value of the strong coupling \as\ obtained from the four-jet rate 
    represents our main result. The analyses of the $D$-parameter and \tmi\ 
    return larger uncertainties as discussed in Section~\ref{eventshapes}.
    Results published by the \Opal\ collaboration using different
    predictions and event shapes are shown as well. The value
    of \as\ indicated by Jet Rates is obtained by ~\cite{prmike}.
    The results of~\cite{pr404} are indicated as 'Event Shape  
    Distributions' and 'Event Shape Moments'.
    The type of QCD prediction is indicated in the plot.
    The \as\ world average is taken from~\cite{alphasbethke}.
  }
  \label{comp_as}
\end{figure}

From a fit of matched $\cal{O}(\ascu)$+NLLA predictions to the four-jet rate,
we have determined the strong coupling \restot, which is consistent with 
the world average value of $\asmz=0.1182\pm0.0027$
~\cite{alphasbethke}, based on a detailed evaluation of systematic errors.
A fit to the $D$-parameter and the \tmi\ was performed using
basic $\cal{O}(\ascu)$ predictions. We find consistent results 
with significantly larger theoretical uncertainties. Figure~\ref{comp_as} 
summarizes the results obtained and compares them with results 
from previous \Opal\ publications using jet-rates~\cite{prmike}, event
shape distributions and moments~\cite{pr404}. The results
are in good agreement with each other. Our result using
the four-jet rate has the smallest uncertainty.
This is mostly due to the comparatively small scale uncertainty,
originating from the fact that the natural scale \xmu=1 is close
to the scale with the least sensitivity to \as, as shown 
in Figure~\ref{xmuopt}.
\appendix
\par
\section*{Acknowledgements:}
\par
We particularly wish to thank the SL Division for the efficient operation
of the LEP accelerator at all energies
 and for their close cooperation with
our experimental group.  In addition to the support staff at our own
institutions we are pleased to acknowledge the  \\
Department of Energy, USA, \\
National Science Foundation, USA, \\
Particle Physics and Astronomy Research Council, UK, \\
Natural Sciences and Engineering Research Council, Canada, \\
Israel Science Foundation, administered by the Israel
Academy of Science and Humanities, \\
Benoziyo Center for High Energy Physics,\\
Japanese Ministry of Education, Culture, Sports, Science and
Technology (MEXT) and a grant under the MEXT International
Science Research Program,\\
Japanese Society for the Promotion of Science (JSPS),\\
German Israeli Bi-national Science Foundation (GIF), \\
Bundesministerium f\"ur Bildung und Forschung, Germany, \\
National Research Council of Canada, \\
Hungarian Foundation for Scientific Research, OTKA T-038240, 
and T-042864,\\
The NWO/NATO Fund for Scientific Research, the Netherlands.\\

\end{document}